\documentclass[5p,twocolumn,10pt,times]{elsarticle}

\usepackage{hyperref}
\usepackage{enumitem}
\usepackage{graphicx}%
\usepackage{multirow}%
\usepackage{amsmath,amssymb,amsfonts}%
\usepackage{amsthm}%
\usepackage{mathrsfs}%
\usepackage[title]{appendix}%
\usepackage{xcolor}%
\usepackage{textcomp}%
\usepackage{manyfoot}%
\usepackage{booktabs}%
\usepackage{algorithm}%
\usepackage{algpseudocode}%
\usepackage{listings}%
\usepackage{subfig}
\usepackage{tikz}
\usepackage{pdflscape}
\usepackage{algorithmicx}

\newtheorem{remark}{Remark}
\begin{document}

\begin{frontmatter}



  \title{DDPM-Polycube: A Denoising Diffusion Probabilistic Model for
    Polycube-Based Hexahedral Mesh Generation and Volumetric Spline
    Construction} %% Article title


\author[1]{Yuxuan Yu*}
\author[1]{Yuzhuo Fang}
\author[2]{Hua Tong}
\author[1]{Jiashuo Liu}
\author[2]{Yongjie Jessica Zhang*}

\affiliation[1]{organization={Institute of Artificial Intelligence, Donghua University},%Department and Organization
  addressline={2999 North Renmin Road},
  city={Shanghai},
  postcode={201620},
  country={China}}

\affiliation[2]{organization={Department of Mechanical Engineering, Carnegie Mellon University},%Department and Organization
  addressline={5000 Forbes Ave},
  city={Pittsburgh},
  postcode={15213},
  state={PA},
  country={USA}}

%% Abstract
\begin{abstract}

  In this paper, we propose DDPM-Polycube, a generative polycube creation
  approach based on denoising diffusion probabilistic models (DDPM) for
  generating high-quality hexahedral (hex) meshes and constructing volumetric
  splines. Unlike DL-Polycube methods that rely on predefined polycube structure
  templates, DDPM-Polycube models the deformation from input geometry to its
  corresponding polycube structures as a denoising task.  By learning the
  deformation characteristics of simple geometric primitives (a cube and a cube
  with a hole), the DDPM-Polycube model progressively reconstructs polycube
  structures from input geometry by removing non-standard Gaussian noise. Once
  valid polycube structures are generated, they are used for surface
  segmentation and parametric mapping to generate high-quality hex
  meshes. Truncated hierarchical B-splines are then applied to construct
  volumetric splines that satisfy the requirements of isogeometric analysis
  (IGA). Experimental results demonstrate that DDPM-Polycube model can directly
  generate polycube structures from input geometries, even when the topology of
  these geometries falls outside its trained range. This provides greater
  generalization and adaptability for diverse engineering geometries. Overall,
  this research shows the potential of diffusion models in advancing mesh
  generation and IGA applications.

\end{abstract}


\end{frontmatter}



\section{Introduction}

Isogeometric analysis (IGA)~\cite{Hughes05a} has undergone substantial
development over the past twenty years. However, the construction of volumetric
parameterization remains an ongoing challenge. The approaches for construction
of volumetric parameterization in IGA can be divided into two primary categories
based on the input models: constructive solid geometry
(CSG)~\cite{zuo_isogeometric_2015} and boundary representation
(B-Rep)~\cite{li_generalized_2010,zhang_solid_2012}. However, CSG-based models
pose difficulties for IGA due to the presence of trimming
surfaces~\cite{zuo_isogeometric_2015}. B-Rep models require the generation of
the control meshes and the construction of volumetric spline basis functions on
the control meshes. In finite element analysis, B-Rep models are typically
discretized into tetrahedral or hexahedral (hex) meshes. While tetrahedral
meshes are widely used in industry due to the availability of multiple automatic
generation strategies. Hex meshes are preferred for their advantages. These
include requiring fewer elements to achieve the same level of
accuracy~\cite{benzley1995comparison}, avoiding the locking issues associated
with tetrahedral meshes~\cite{francu2021locking}, and providing better
compatibility with tensor-product spline construction.

While there has been considerable progress in hex mesh
generation~\cite{pietroni_hex-mesh_2023,ref:zhangbook,zhang2013challenges}, it
is challenging to generate high-quality hex meshes for complex B-Rep models.
Various methods have been explored, including indirect methods~\cite{E96},
sweeping methods~\cite{Zhang20072943,yu2019anatomically}, grid-based
methods~\cite{qian2012automatic,schneiders1996grid,qian2010quality}, polycube
methods~\cite{Tarini2004,Wang07polycubesplines,LeiLiu2012a,HZ2015CMAME,yu2020hexgen},
and vector field-based methods~\cite{nieser2011cubecover,Li2012AMU}.  However,
not all of these methods are suitable for generating control meshes for IGA,
which are used for constructing splines. Hex meshes often contain extraordinary
edges and extraordinary points. When extraordinary points and extraordinary
edges are involved, achieving optimal convergence rates becomes a challenging
task in IGA. Therefore, among the various available methods, those that can
generate meshes with as few extraordinary points and extraordinary edges as
possible are preferred. In this context, sweeping and polycube methods are
preferred for generating hex control meshes for IGA. Sweeping methods generate
hex meshes by scanning from source to target surfaces, but their applicability
is limited to models with compatible topologies between the source and target
surfaces. Polycube methods generate hex meshes by using polycube structures as
parameter spaces, along with parametric mapping. The methods were initially used
as a texture mapping technique~\cite{Tarini2004}. Lin \textit{et
  al}.~\cite{LinJ2008} developed an automated method for constructing polycubes;
however, this method is not well-suited for models with complex topology. One of
the key advantages of the polycube method is its ability to control the number
and placement of extraordinary points and edges. As a result, this concept has
been further extended for use in hex mesh generation. Hu \textit{et
  al}.~\cite{HZ2015CMAME,HZL2016,yu2020hexgen} employed centroidal Voronoi
tessellation to segment surfaces. The structure of the polycube can be defined
using the segmentation information, which is then used to generate hex control
meshes through parametric mapping. Guo \textit{et
  al}.~\cite{guo_cut-enhanced_2020} modified the polycube structure by
introducing extraordinary edges to enhance mesh quality. Li \textit{et
  al}.~\cite{LiBo2012} introduced the generalized polycube method, significantly
expanding the adaptability of polycube techniques to high-genus and complex
B-Rep models.

Constructing splines on hex control meshes is another challenge. Techniques like
NURBS~\cite{Zhang20072943},
T-splines~\cite{Wang07polycubesplines,zhang_solid_2012,LeiLiu2012a}, and
TH-splines~\cite{wei17a} have been developed for volumetric parameterization.
T-splines support highly localized refinement. The truncation mechanism in
TH-splines helps reduce the overlap of basis functions from different levels,
thereby improving numerical conditioning. In addition, other researchers have
also explored advancements in local refinement. For instance, Xu \textit{et
  al}.~\cite{xu2019efficient} optimized control point positions in regions with
significant geometric features using an r-adaptive framework. Li \textit{et
  al}.~\cite{LiBo2012} used the local refinement capability of T-splines to
reduce geometric errors in surface fitting.

Recent efforts in artificial intelligence have affect mesh
generation. Mesh\-GPT~\cite{Siddiqui2024} uses a decoder-only transformer to
generate triangular meshes. For tetrahedral mesh generation, the
DefTet~\cite{Gao2020} achieves high-quality tetrahedral reconstruction through
optimization of vertex distribution and volumetric occupancy. The
MeshAnything~\cite{Chen2024} framework generates triangular meshes from point
clouds. In addition to triangular and tetrahedral meshes, Tong \textit{et
  al}.~\cite{TONG2023102109} integrates the advancing front method with neural
networks to generate planar unstructured quadrilateral meshes. For hex mesh
generation, Yu \textit{et al}.~\cite{yu2024dlpolycubedeeplearningenhanced}
integrates deep learning with the polycube method to generate hex control meshes
for IGA.  By utilizing a deep neural network, the algorithm directly predicts
polycube structures, eliminating the need for traditional heuristic adjustments
and additional post-processing steps. However, to enable the model to predict
the ideal polycube structures, the algorithm requires as much training data as
possible to cover the variability of design parameters and topologies.
Generating all possible polycube structures would be an enormous task.
Fortunately, polycube structures have a key characteristic: they can be
decomposed into genus-0 and genus-1 geometries. Therefore, using generative
models appears to be a feasible approach. By training generative models, it is
possible to generate these two primitive geometries and then combine them into
more complex polycube structures. This approach has the potential to improve
efficiency and enhance the model's generalization ability, enabling it to better
predict and generate ideal polycube structures.

Generative models can be categorized into five main types. The first type is
sequence-to-sequence models~\cite{sutskever2014sequence}, which belong to
autoregressive decoders. These models generate sequence elements step by step
and are widely used in natural language processing and time-series generation
tasks. The second type is generative adversarial network models
(GAN)~\cite{goodfellow2014generative}, which do not explicitly model the target
distribution. Instead, through adversarial training between the generator and
discriminator, GANs enable the generator to produce samples that approximate the
target distribution. The third type is flow-based
models~\cite{dinh2016density,kingma2018glow}, which follow a fully reversible
generative process and allow direct modeling of data distributions. However,
their performance is limited by the need for carefully designed structures. The
fourth type is variational autoencoders (VAE)~\cite{kingma2013auto}, which
introduce latent variables and optimize reconstruction errors, enabling the
generation of diverse and continuous samples. The fifth type is diffusion
models~\cite{sohl2015deep,ho2020denoising}, which share certain similarities
with VAEs. The core idea of diffusion models is to learn the data distribution
through a process of gradually adding and removing noise. Diffusion models have
demonstrated significant advantages in various generative tasks in recent
years. Ho~\textit{et al}.~\cite{ho2020denoising} proposed denoising diffusion
probabilistic models, which iteratively learn the reverse diffusion process to
generate high-quality images. These models outperform
GANs~\cite{goodfellow2014generative} and VAEs~\cite{kingma2013auto} in terms of
sample quality and stability. Unlike the adversarial training of GANs, diffusion
models adopt maximum likelihood estimation, ensuring the stability of the
training process and avoiding issues such as mode collapse. Song~\textit{et
  al}.~\cite{song2020score} further introduced a score-based generative
framework. This model improves the stability of training by mapping the
generation process to stochastic differential equations. Nichol~\textit{et
  al}.~\cite{nichol2021improved} showed that by gradually denoising, diffusion
models can effectively reduce the interference of noise and outliers during the
generation process, thereby improving the quality of model generation.

In this paper, we introduce a novel method to generate high-quality hex meshes
and construct volumetric splines by leveraging denoising diffusion probabilistic
model (DDPM) and polycube techniques. Building upon the foundation of DDPM, we
present the DDPM-Polycube algorithm, which utilizes diffusion models to generate
polycube structures. Then the method segments geometry surfaces to match the
polycube structure and generate hex meshes through parametric mapping.  In
summary, this paper presents four main contributions:

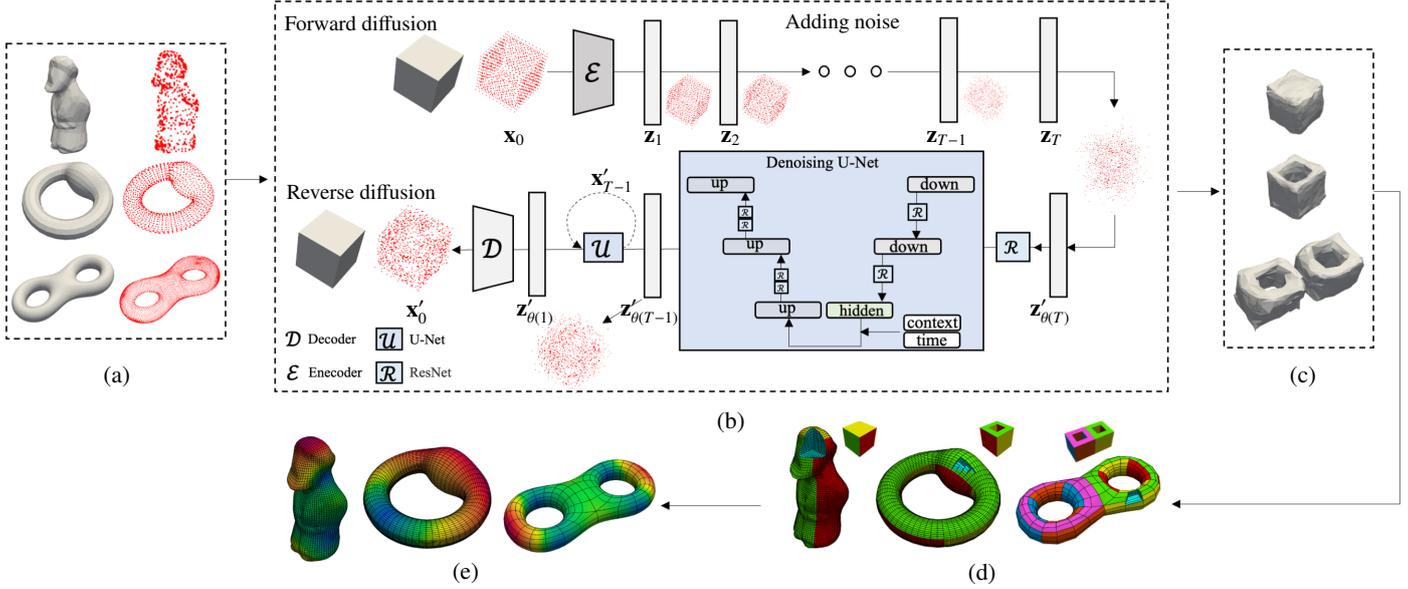
\begin{figure*}[!ht]
      \centering
  \begin{tikzpicture}
    \node[anchor=south west,inner sep=0] (image) at (0,0) {\includegraphics[width=\linewidth]{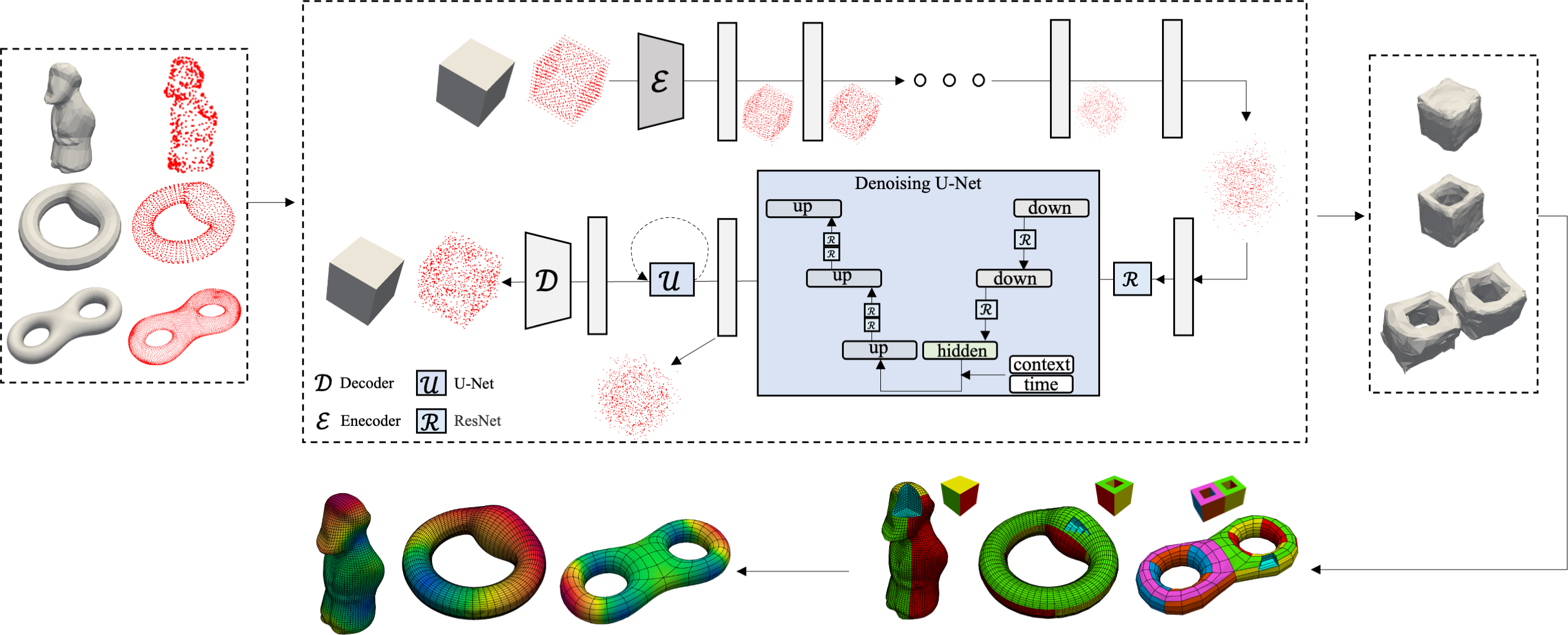}};
    \begin{scope}[x={(image.south east)},y={(image.north west)}]
      \node at (0.255,0.96) { \footnotesize Forward diffusion};
      \node at (0.255,0.66) { \footnotesize Reverse diffusion};
      \node at (0.60,0.96) { \footnotesize Adding noise};
      \node at (0.365,0.755) {\small \(\mathbf{x}_0\)};
      \node at (0.465,0.755) {\small \(\mathbf{z}_1\)};
      \node at (0.52,0.755) {\small \(\mathbf{z}_2\)};
      \node at (0.675,0.755) {\small \(\mathbf{z}_{T-1}\)};
      \node at (0.75,0.755) {\small \(\mathbf{z}_{T}\)};
      \node at (0.295,0.445) {\small \(\mathbf{x}'_0\)};
      \node at (0.38,0.445) {\small \(\mathbf{z}'_{\theta(1)}\)};
      \node at (0.462,0.445) {\small \(\mathbf{z}'_{\theta (T-1)}\)};
      \node at (0.749,0.445) {\small \(\mathbf{z}'_{\theta (T)}\)};
      \node at (0.435,0.68) {\small \(\mathbf{x}'_{T-1}\)};
      % % $\mathcal{D}$ Decoder $\square$ U U-Net $\mathcal{U}$
      % % $\mathcal{E}$ Enecoder $\square$ $\mathcal{R}$ ResNet
      % \node at (0.38,0.655) {\scriptsize Forward diffusion};
      % \node at (0.59,0.655) {\scriptsize Reverse diffusion};
      % \node at (0.48,0.3) {\scriptsize Applications};
      % \node at (0.155,0.2) {\scriptsize Denoise};
      % \node at (0.48,0.2) {\scriptsize Denoise};
      % \node at (0.78,0.2) {\scriptsize Denoise};
      \node at (0.08,0.33) {\small (a)};
      \node at (0.52,0.25) {\small (b)};
      \node at (0.93,0.33) {\small (c)};
      \node at (0.7,-0.02) {\small (d)};
      \node at (0.33,-0.02) {\small (e)};
    \end{scope}
  \end{tikzpicture}
  \vspace{-2em}\caption{\label{fig:Pipeline_Structure}{The DDPM-Polycube
      pipeline. (a) Converting CAD geometries into triangular meshes and point
      clouds; (b) the DDPM-Polycube model; (c) the polycube structure generated
      by the DDPM-Polycube model; (d) an all-hex control mesh generated through
      octree subdivision, parametric mapping and quality improvement techniques,
      some elements are removed to show the interior; and (e)
      volumetric spline with IGA simulation results using ANSYS-DYNA. The
      diffusion model relies on two substeps: (1) a forward diffusion process
      that adds non-standard Gaussian noise to the point cloud data of the
      polycube structure, and (2) a reverse diffusion process that removes the
      noise to reconstruct the polycube structure.}}
\end{figure*}

\begin{enumerate}
\item We propose an innovative perspective that the deformation from input
  geometry to polycube structures can be viewed as a denoising task. In this
  context, any input geometries are seen as polycube structures with many
  small-scale deformations superimposed. These deformations follow a
  non-standard Gaussian distribution. The algorithm employs a parameterized
  Markov chain to reverse the diffusion process, deforming the input geometry to
  polycube structures. The polycube structure generated by this process is then
  used for hex mesh generation and volumetric spline construction.

\item We have proposed a novel diffusion and reverse diffusion process equation
  specifically designed for hex mesh generation and volumetric spline
  construction. Our model is unique in its ability to handle non-standard
  Gaussian distribution, distinguishing it from many traditional diffusion
  models. This requires modifying the model to incorporate additional parameters
  and adjustments. By leveraging these enhancements, our method ensures that the
  reverse diffusion process generate the polycube structures. This innovation
  broadens the application scope of diffusion models in computational geometry
  and computational mechanics.

\item The DDPM-Polycube method differs from the DL-Polycube
  classification~\cite{yu2024dlpolycubedeeplearningenhanced} by using diffusion
  models to construct polycube structures directly. Unlike the DL-polycube
  methods, which focus on learning mappings and are restricted by the ground
  truth structures in the dataset, the DDPM-Polycube method focuses on learning
  the deformation, thereby improving the model's generalization ability. It is
  particularly effective considering the wide variety of polycube structures
  found in real-world engineering geometries.

\item Unlike traditional polycube methods, DDPM-polycube does not require
  additional post-processing steps and heuristic operations to ensure valid
  structures~\cite{pietroni_hex-mesh_2023}, as our DDPM-polycube algorithm
  leverages a diffusion model to directly generate polycube structures. Then, it
  performs surface segmentation (labeling) based on these predictions. Since the
  segmentation or labeling is inherently based on polycube structures, it
  naturally guarantees valid polycube structures.

\end{enumerate}

The remainder of this paper is organized as follows. Section 2 provides an
overview of the DDPM-Polycube algorithm design and pipeline. Section 3 details
dataset generation, feature extraction and the diffusion model
architecture. Section 4 discusses how polycube structures generated using
diffusion models are utilized for high-quality hex mesh generation and
volumetric spline construction. Section 5 presents examples that demonstrate the
algorithm's efficiency and effectiveness. Finally, Section 6 concludes the paper
and suggests directions for future research.

\allowdisplaybreaks
\section{Overview of the pipeline}

As shown in Fig.~\ref{fig:Pipeline_Structure}, our DDPM-Polycube pipeline begins
with a triangular mesh representing the CAD geometry. Subsequently, our
DDPM-Polycube model is employed to directly generate polycube structures. This
polycube structure serves as the foundation for generating all-hex meshes
through surface segmentation and parametric mapping~\cite{floater1997}.  To
ensure that the generated all-hex mesh meets the quality requirements necessary
for IGA, we employ several mesh quality improvement techniques—such as
pillowing~\cite{YZhang2009c}, smoothing, and
optimization~\cite{zhang_solid_2012, qian2012automatic,
  tong_hybridoctree_hex_2024}—as needed.  Upon achieving a high-quality hex
mesh, the volumetric spline model is constructed from these meshes using
TH-spline3D with local refinement~\cite{wei17a,yu2020hexgen}. Finally, the
B\'{e}zier information is extracted for performing IGA in ANSYS-DYNA.

\section{DDPM-Polycube model}

As shown in Fig.~\ref{fig:Pipeline_Structure}, we propose a new DDPM-Polycube
approach, which iteratively deforms the input geometry into corresponding
polycube structures. Unlike DL-Polycube
methods~\cite{yu2024dlpolycubedeeplearningenhanced}, which rely on the
predefined polycube types and mapping between the input geometry and output
polycube types. The DDPM-Polycube algorithm models the geometric deformation
process as a denoising task.

In the preprocessing step, the input geometries are first converted into point
cloud data, which then serves as the input to the DDPM-Polycube model. The
DDPM-Polycube model operates in two steps (see
Fig.~\ref{fig:Pipeline_Structure}): (1) a forward diffusion process, where
non-standard Gaussian noise is incrementally added to the point cloud data, and
(2) a reverse diffusion process, where the model progressively removes the
deformation to recover the target polycube structure. During the training step,
the model learns the deformation capability from simple geometric primitives: a
cube and a cube with a hole. Once training is complete, the model generalizes to
more complex geometries. The DDPM-Polycube algorithm consists of data
generation, feature extraction, and training the improved DDPM model. These
steps work together to generate polycube structures from input geometries.

% ~\cite{HZ2015CMAME,HZL2016,Liu2015,guo_cut-enhanced_2020,LeiLiu2012a,Wenyan2013c,zhang_solid_2012}.

\subsection{Data generation}
\label{sec:data-generation}

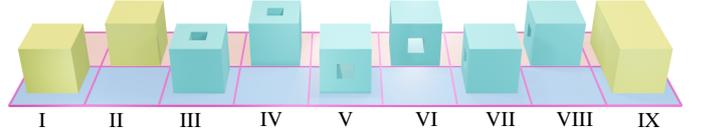
\begin{figure}[t]
      \centering
  \begin{tikzpicture}
    \node[anchor=south west,inner sep=0] (image) at (0,0) {\includegraphics[width=\linewidth]{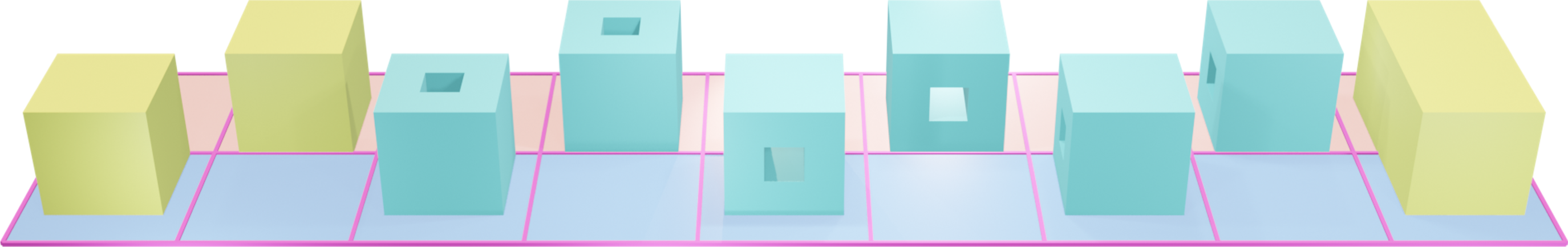}};
    \begin{scope}[x={(image.south east)},y={(image.north west)}]
      \node at (0.05,-0.12) {\footnotesize I};
      \node at (0.16,-0.12) {\footnotesize II};
      \node at (0.27,-0.12) {\footnotesize III};
      \node at (0.39,-0.12) {\footnotesize IV};
      \node at (0.5,-0.12) {\footnotesize V};
      \node at (0.62,-0.12) {\footnotesize VI};
      \node at (0.73,-0.12) {\footnotesize VII};
      \node at (0.84,-0.12) {\footnotesize VIII};
      \node at (0.95,-0.12) {\footnotesize IX};
    \end{scope}
  \end{tikzpicture}
  \vspace{-2em}\caption{\label{fig:05_context}{9 geometric configuration types
      for dataset generation. The first 8 columns contain one geometric
      primitive, either a cube or a cube with a hole oriented along the $Z$, $X$, or
      $Y$ axis, placed in different positions. The 9th column represents composite
      structures formed by two primitives. These configurations, along with the
      one-hot encoding, provide the starting point for the forward diffusion
      process.  }}
\end{figure}

In the data generation step, we focus on creating a context-rich dataset
starting from two basic geometric primitives: a cube (genus-0) and a cube with a
hole (genus-1). These models are generated and discretized into triangular
meshes using the 3D graphics software Blender, along with its built-in Python
and BMesh libraries. To provide additional contextual information, we further
assign these geometric primitives to specific positions and treat geometric
primitives at different positions as different types. As shown in
Fig.~\ref{fig:05_context}, the first 8 columns are composed of single geometric
primitive, while the 9th column contains a combination of two geometric
primitives. Note that the cube with a hole is further categorized based on the
axis along which the hole is oriented ($X$-axis, $Y$-axis, or $Z$-axis),
resulting in two additional variations. To assign the positions of these
geometric primitives, we design a $2\times 1$ grid \( G_{2\times 1} \). The
\( G_{2\times 1} \) grid consists of two units arranged in one column with two
rows (see Fig.~\ref{fig:05_context} for an illustration of the grid for each
type of geometric configurations).

To enable the DDPM-Polycube model to learn more complex structures, we introduce
a 9th type, represented by the 9th column shown in
Fig.~\ref{fig:05_context}. This type consists of two cubes and occupy two units
of the \( G_{2\times 1} \) grid. One cube is placed on the top of the other. The
inclusion of this composite geometric configuration type is essential to help
the DDPM-Polycube model understand that polycube structures can be formed by
combining multiple geometric primitives. Our experiments show that without this
9th type, the model fails to learn the ability to combine geometric primitives
effectively. As a result, it can only generate a single cube or a cube with a
hole, and fails to generate more complex polycube structures composed of
multiple geometric primitives.

To contextualize these types, we use a one-hot encoding scheme, a common method
for representing discrete features. This encoding assigns a unique binary vector
to each type, where the vector’s length is 29, and each bit indicates the
presence or absence of a specific geometric type or combination. For geometric
primitives types (the first 8 types), each type occupies a distinct position in
the one-hot encoding vector, and the corresponding bit is set to 1 to indicate
its presence. For the 9th composite type, two positions in the vector are
activated simultaneously to indicate the presence of two geometric
primitives. The encoding rules are as follows: for the first 8 types,
$\text{context}=\delta_{n, 4\mathcal{T}}, \quad t = 0, 1, \dots, 7$; for the 9th type,
$\text{context}=\delta_{n, 0}+\delta_{n, 4}$, where \( n \) represents the
position, \( \delta \) is the Kronecker delta, and \(\mathcal{T} \) represents the
geometric type.  Through this encoding method, each type of geometric
configuration is uniquely represented, enabling the model to recognize and learn
from them effectively.

With these 9 types of geometric configurations, we can generate a context-aware
dataset. For each of those 9 types of geometric configurations, we introduce
non-standard Gaussian noise to these geometries. This noise can be interpreted
as small-scale deformations that progressively obscure the original polycube
structure. By applying this noise, we generate a series of deformed geometry,
forming a class of latent variable models.  Finally, the dataset is used for the
reverse diffusion process, enabling the model to learn the ability to generate
polycube structures by gradually deforming (denoising) the input geometry.

\subsection{Feature extraction}

The feature extraction process begins by converting the triangular mesh into a
point cloud, denoted as
$ P = \{ \mathbf{p}_1, \mathbf{p}_2, \dots, \mathbf{p}_N \} $, where each point
$ \mathbf{p}_i = \mathbf{x}_i $ represents the 3D coordinates of the $ i $-th
point, with $ \mathbf{x}_i = (x_i, y_i, z_i) $. Here, $ N $ is the total number
of points. The 3D coordinates of the point cloud are then normalized to the
range $[-1, 1]$. After normalization, the coordinates are further scaled to the
range $[-0.5, 0.5]$ and then shifted by 0.5 along the $X$-axis. Subsequently,
all data are sorted along the $X$-axis. For the first eight types of geometric
configurations (the first eight columns shown in Fig.~\ref{fig:05_context}),
each type occupies one unit in the \( G_{2\times 1} \) grid and contains 512
points. To ensure consistent data size across all geometric configurations, an
additional 512 zeros are appended to these 512 points, resulting in a total of
1,024 points for each geometric configuration. For polycube structure that
occupy two units in the \( G_{2\times 1} \) grid (the 9th column shown in
Fig.~\ref{fig:05_context}), it contains 1,024 points without the need for zero
padding. Regardless of the number of units occupied in the \( G_{2\times 1} \)
grid, the point clouds of all geometric configurations are processed and
organized into a three-channel \( 32 \times 32 \) matrix. This process maps the
points of the triangular mesh into a $32\times 32$ matrix, creating images that
serve as inputs for the DDPM-Polycube model. By leveraging image-like data
structures, this method aligns with the DDPM framework.

During the reverse diffusion process, the input geometry is considered as $x'_T$ and
requires normalization to ensure consistency with the dataset's scale. It is
then converted into a three-channel $32\times 32$ matrix using the same data
structure of the training data, serving as the input for the reverse diffusion
process. Then, the result generated by the reverse diffusion process is in
three-channel $32\times 32$ matrix. These need to be decoded back into 3D
coordinates. Finally, the generated point cloud data maintains the same number
of points as the vertices of the original triangular mesh and has the same
topology.

\begin{remark}[Training data design]
  The training data for the DDPM-Polycube algorithm includes nine types, based
  on two geometric primitives (a cube and a cube with a hole) and a combination
  of two cubes. While these two geometric primitives form the foundation, other
  basic geometric primitives can also be included. The combination configuration
  helps the model learn how geometric primitives merge, enabling it to
  reconstruct more complex structures.
\end{remark}

\begin{remark}[Grid configuration design]
  This study uses a simplified $2\times 1$ grid configuration as a
  proof-of-concept to reduce complexity. Despite the limited dataset, the
  results show the model can generalize effectively, deforming complex CAD
  geometries into polycube structures, even for topologies beyond its training
  range.  The algorithm is also extensible, supporting larger grid
  configurations such as a \(3 \times 2\) grid, \(m \times n\) grids, or even
  higher-dimensional grids.  This would enable it to handle more complex
  geometries and topologies.
\end{remark}

\subsection{DDPM-Polycube model architecture}
\label{sec:diffusion-framework}

The improved DDPM plays a key role in DDPM-Polycube model, enabling the
deformation and reconstruction of corresponding polycube structures. It operates
through two diffusion processes: the forward diffusion process and the reverse
diffusion process. In the forward diffusion process, non-standard Gaussian noise
is progressively added to the polycube structure (as shown in
Fig.~\ref{fig:Pipeline_Structure}), gradually deforming it into a geometry with
a Gaussian noise. The reverse diffusion process iteratively removes the
deformation, reconstructing the corresponding polycube structure step by step
through deformation (denoising).

\subsubsection{Forward diffusion process}
\label{sec:forw-diff-proc}

The forward diffusion process is a Markov chain that progressively corrupts the
polycube structure \(\mathbf{x}_0\) by introducing non-standard Gaussian noise
at each timestep \(t\). Using the reparameterization trick, the non-standard
Gaussian noise \(\mathcal{N}(\mathbf{q}, I)\) is reformulated as the sum of
standard Gaussian noise \(\mathcal{N}(0, I)\) and the adjustment term
\(\mathbf{q}\). Consequently, the geometry at timestep \(t\) is expressed as:
\begin{equation}
  \label{eq:1}
  \mathbf{x}_t = \sqrt{\alpha_t} \mathbf{x}_{t-1} + \sqrt{1-\alpha_t} (\mathbf{z}_{t-1} + \mathbf{q}), \quad \mathbf{z}_t \sim \mathcal{N}(0, I),
\end{equation}
where \(\mathbf{x}_t\) represents the deformed geometry at timestep \(t\),
\(\alpha_t = 1 - \beta_t\), and \(\beta_t\) is a predefined deformation
schedule. The term \(\mathbf{z}_t\) denotes standard Gaussian noise, and
\(\mathbf{q}\) is a new term introduced in this paper that differentiates this
method from the standard DDPM framework~\cite{ho2020denoising}.

To further simplify the forward diffusion process, we reformulate
Equation~\eqref{eq:1} in terms of \(\mathbf{x}_0\) and \(\mathbf{q}\). This
reformulation enables the computation of \(\mathbf{x}_t\) for any timestep \(t\)
without requiring intermediate iterations or latent variable models. By
leveraging the reparameterization trick, the forward diffusion process can be
expressed as:
\begin{equation}
\label{eq:2}
\begin{aligned}
  \mathbf{x}_t & =\sqrt{\alpha_t} \mathbf{x}_{t-1}+\sqrt{1-\alpha_t}\left(\mathbf{z}_{t-1}+\mathbf{q}\right) \\
               & =\sqrt{\alpha_t}\left(\sqrt{\alpha_{t-1}}
                 \mathbf{x}_{t-2}+\sqrt{1-\alpha_{t-1}}\left(\mathbf{z}_{t-2}+\mathbf{q}\right)\right)\\
               &\hspace{5mm}+\sqrt{1-\alpha_t}\left(\mathbf{z}_{t-1}+\mathbf{q}\right) \\
      % & =\sqrt{\bar{\alpha}_t} \mathbf{x}_0+\sqrt{1-\bar{\alpha}_t} \bar{\mathbf{z}}_t+\left(\sum_{k=1}^t \sqrt{\left(1-\alpha_k\right) \prod_{i=k+1}^t} \alpha_i\right) \mathbf{q} \\
      & =\sqrt{\bar{\alpha}_t} \mathbf{x}_0+\sqrt{1-\bar{\alpha}_t} \bar{\mathbf{z}}_t+\mathbf{Q}_t,
\end{aligned}
\end{equation}
where $\mathbf{z}_{t-1}, \mathbf{z}_{t-2}, \ldots, \sim \mathcal{N}(0, I)$,
\(\bar{\alpha}_t = \prod_{i=1}^t \alpha_i\) is the cumulative deformation decay,
\(\bar{\mathbf{z}}_t\) aggregates the Gaussian distribution up to timestep
\(t\), and
\(\mathbf{Q}_t=\left(\sum_{k=1}^t \sqrt{\left(1-\alpha_k\right) \prod_{i=k+1}^t}
  \alpha_i\right) \mathbf{q}\) is a term that accounts for the cumulative
influence of \(\mathbf{q}\) across all previous timesteps. From
Equation~\eqref{eq:2}, the conditional probability
\(q(\mathbf{x}_t | \mathbf{x}_0)\) can be derived as:
\begin{equation}
  q(\mathbf{x}_t | \mathbf{x}_0) = \mathcal{N}(\mathbf{x}_t; \sqrt{\bar{\alpha}_t} \mathbf{x}_0+\mathbf{Q}_t, (1-\bar{\alpha}_t)I),
\end{equation}
where the mean is scaled by \(\sqrt{\bar{\alpha}_t}\), and the variance is
determined by \(1-\bar{\alpha}_t\), reflecting the cumulative deformation
applied up to timestep \(t\). The inclusion of \(\mathbf{q}\) in
\(\mathbf{Q}_t\) adjusts the forward diffusion process to account for the
relationship between the endpoint of the forward diffusion process and
the polycube structure.
\begin{remark}[Non-standard Gaussian noise]
  In the standard DDPM framework~\cite{ho2020denoising}, when the timestep \(t\)
  approaches infinity, the data converges into a standard Gaussian
  distribution. However, this does not meet the requirements of our method. For
  the DDPM-Polycube algorithm, the endpoint of the forward diffusion process
  (\(\mathbf{x}_T\)) needs to be represented as a geometry. To achieve this, we
  introduce a non-standard Gaussian noise, where \(\mathbf{q}\) is defined as an
  incremental adjustment term applied during the forward diffusion process. This
  adjustment term deforms the polycube structure (\(\mathbf{x}_0\)) into the
  final geometry (\(\mathbf{x}_T\)).
\end{remark}
\begin{remark}[Computation of $\mathbf{q}$]
  When the timestep \(t\) approaches infinity, \(\mathbf{x}_t\) converges to
  \(\bar{\mathbf{z}}_t + \mathbf{Q}_t\), as described in
  Equation~\eqref{eq:2}. Here, \(\mathbf{x}_t\) represents the endpoint of the
  forward diffusion process, which also serves as the initial state (input
  geometry) for the reverse diffusion process. Then, the adjustment term
  \(\mathbf{Q}_t\) can be calculated. Since
  \(\mathbf{Q}_t=\left(\sum_{k=1}^t \sqrt{\left(1-\alpha_k\right)
      \prod_{i=k+1}^t} \alpha_i\right) \mathbf{q}\), the value of \(\mathbf{q}\) can be directly derived.
\end{remark}

\subsubsection{Reverse diffusion process}
\label{sec:reverse-diff-proc}
The reverse diffusion process aims to reconstruct the input geometry
\(\mathbf{x}'_t\) to the corresponding polycube structure
(\(\mathbf{x}'_0\)). By utilizing a parameterized neural network
\(\mathbf{z}'_\theta(\mathbf{x}'_t, t)\), \(\mathbf{x}'_t\) is iteratively
deformed. This ensures that the original geometric topology is preserved while
gradually generating the corresponding polycube structure. Mathematically, the
reverse diffusion process is modeled as a conditional probability distribution:
\setlength{\abovedisplayskip}{3pt}
\begin{equation}
  \label{eq:3}
  q(\mathbf{x}'_{t-1} | \mathbf{x}'_t, \mathbf{x}'_0) = \mathcal{N} \left(\mathbf{x}'_{t-1}; \mu_t'(\mathbf{x}'_t, \mathbf{x}'_0), \sigma_t^2 I \right),
\end{equation}
where \(\mu_t'(\mathbf{x}'_t, \mathbf{x}'_0)\) is the mean of the corresponding
Gaussian distribution, which represents a weighted combination of
\(\mathbf{x}'_t\) and \(\mathbf{x}'_0\), and
\(\sigma_t^2 = \frac{1 - \bar{\alpha}_{t-1}}{1 - \bar{\alpha}_t} \beta_t\) is
the variance at timestep \(t\). By iteratively estimating this conditional
distribution, the model effectively reconstructs the geometry generated during
the forward process.

To compute \(\mathbf{x}'_{t-1}\), solving for the mean \(\mu_t'\) is the core
step in the reverse diffusion process, as it guides the reconstruction of
geometry at each timestep \(t\). Referring to the DDPM~\cite{ho2020denoising}
and adapting it for our polycube reconstruction task, we can get similar
function:
\begin{equation}
  \label{eq:5}
  \mathbf{\mu}_{t}' = \frac{1}{\sqrt{\alpha_t}} \left( \mathbf{x}'_t - \frac{\beta_t}{\sqrt{1-\bar{\alpha}_t}} (\mathbf{z}'_\theta(\mathbf{x}'_t, t) +\mathbf{q}')\right).
\end{equation}
The parameter \(\mathbf{q}'\) is an adjustment term introduced to ensure the
integrity of the reconstructed polycube structure, while the neural network
\(\mathbf{z}'_\theta(\mathbf{x}'_t, t)\) is trained to predict the deformation added
during the forward diffusion process. Compared to the equation in the
literature~\cite{ho2020denoising}, our method introduces an additional parameter
\(\mathbf{q}'\), which can be derived as follows.

To reconstruct \(\mathbf{x}'_{t-1}\), the reverse diffusion process combines the
predicted mean \(\mu_t'\) with the variance term \(\sigma_t \mathbf{z}'_t\),
resulting in the following equation:
\begin{equation}
  \begin{aligned}
  \mathbf{x}'_{t-1}&=\mathbf{\mu}_{t}'+\sigma_t
                     \mathbf{z}_t'\\
    &=\frac{1}{\sqrt{\alpha_t}} \left( \mathbf{x}'_t -
    \frac{\beta_t}{\sqrt{1-\bar{\alpha}_t}} (\mathbf{z}'_\theta(\mathbf{x}'_t,
    t)+\mathbf{q}')\right)+\sigma_t \mathbf{z}_t'.
  \end{aligned}
\end{equation}
By iteratively applying the reverse diffusion process across all timesteps, the
target polycube structure \(\mathbf{x}'_0\) can be progressively
reconstructed. Starting from \(\mathbf{x}'_0\) at timestep \(0\), the process can
be expressed as:
\begin{equation}
\begin{aligned}
  \mathbf{x}_0^{\prime} & =\frac{1}{\sqrt{\alpha_1}}\left(\mathbf{x}_1^{\prime}-\frac{\beta_1}{\sqrt{1-\bar{\alpha}_1}}\left(\mathbf{z}_1+\mathbf{q}^{\prime}\right)\right)+ \sigma_1 \mathbf{z}_1'\\
               % & =\frac{1}{\sqrt{\alpha_1}}\left(\mathbf{x}_1^{\prime}-\frac{\beta_1}{\sqrt{1-\bar{\alpha}_1}} \mathbf{q}^{\prime}\right)+\tilde{\mathbf{z}}_1 \\
               &
                 =\frac{1}{\sqrt{\alpha_1}}\left(\frac{1}{\sqrt{\alpha_2}}\left(\mathbf{x}_2^{\prime}-\frac{\beta_2}{\sqrt{1-\bar{\alpha}_2}} \mathbf{q}^{\prime}\right)\right)\\
                        &\hspace{5mm}-\frac{1}{\sqrt{\alpha_1}} \frac{\beta_1}{\sqrt{1-\bar{\alpha}_1}} \mathbf{q}^{\prime}+\mathbf{z} \\
               &
                 =\frac{1}{\sqrt{\bar{\alpha}_t}}\left(\mathbf{x}_t^{\prime}\right)+\tilde{\mathbf{z}}-\mathbf{Q}',
\end{aligned}
\end{equation}
where
$\mathbf{Q}'=\sum_{k=1}^t \frac{1}{\sqrt{\bar{\alpha}_t}}
\frac{\sqrt{1-\alpha_k}}{\sqrt{1-\bar{\alpha}_k}} \sqrt{1-\alpha_k}
\prod_{i=k+1}^t \sqrt{\alpha_i} \mathbf{q}^{\prime}$ represents the cumulative
deformation adjustments across all timesteps, \(\tilde{\mathbf{z}}\), derived
through the reparameterization trick, aggregates the Gaussian
distribution. Under ideal conditions, when \(t = T\), where
\(\mathbf{x}_t = \mathbf{x}_t'\) and \(\mathbf{x}_0 = \mathbf{x}_0'\), the value
of \(\mathbf{q}'\) can be derived by combining this assumption with
Equation~\eqref{eq:2}. We have:
\begin{equation}
  \begin{aligned}
    \sum_{k=1}^t \frac{1}{\sqrt{\bar{\alpha}_t}} \frac{\sqrt{1-\alpha_k}}{\sqrt{1-\bar{\alpha}_k}} \sqrt{1-\alpha_k} \prod_{i=k+1}^t \sqrt{\alpha_i}\mathbf{q}^{\prime}=\frac{\mathbf{Q}_t}{\sqrt{\bar{\alpha}_t}}. \\
  \end{aligned}
   \end{equation}
Then, we obtain
   \begin{equation}
   \begin{aligned}
     \mathbf{q}^{\prime}=\frac{\sqrt{1-\bar{\alpha}_k}}{\sqrt{1-\alpha_k}}\mathbf{q}.
    \end{aligned}
\end{equation}
This indicates that the value of \(\mathbf{q}'\) depends on the timestep \(k\)
and the diffusion adjustment term \(\mathbf{q}\).

\begin{remark}[Reparameterization trick]
  The reparameterization trick leverages the statistical properties of Gaussian
  distributions. For example, given two independent Gaussian distributions
  \(X \sim \mathcal{N}(\mu_1, \sigma_1^2)\) and
  \(Y \sim \mathcal{N}(\mu_2, \sigma_2^2)\), their linear combination
  \(aX + bY\) results in a new Gaussian distribution with a mean of
  \(a\mu_1 + b\mu_2\) and a variance of \(a^2\sigma_1^2 + b^2\sigma_2^2\).  This
  principle is applied in the DDPM-Polycube algorithm to derive the cumulative
  impact of \(\mathbf{q}\) and \(\mathbf{q}'\) over multiple timesteps during
  the forward diffusion process and the reverse diffusion process.
\end{remark}

\subsubsection{Training the DDPM-Polycube model}
\label{sec:model-training}

The training step of DDPM-Polycube focuses on training the deformation predictor
\(\mathbf{z}'_\theta(\mathbf{x}_t, t)\) to learn the reverse diffusion process,
enabling it to predict the small-scale deformations added during the forward
diffusion process. The loss function is defined as the mean squared error (MSE)
between the true noise \(\mathbf{z}\) and the predicted noise
\(\mathbf{z}'_\theta(\mathbf{x}_t, t)\):
\begin{equation}
  \label{eq:4}
  \mathcal{L} = \mathbb{E}_{\mathbf{x}_0, \mathbf{z}, t} \left[  \| \mathbf{z} - \mathbf{z}'_\theta(\mathbf{x}_t, t) \|^2 \right],
\end{equation}
where \(\mathbf{x}_t\) represents the target geometry generated during the
forward diffusion process according to Equation~\eqref{eq:2}. By iteratively
optimizing this loss function, the neural network learns to predict the
deformation at each timestep, enabling effective deformed during the reverse
diffusion process.

\begin{algorithm}[!htbp]
\caption{Training the DDPM-Polycube model}
\begin{algorithmic}[1]
  \Require Dataset of polycube structures $\{\mathbf{x}_0\}$, number of diffusion steps $T$, deformation schedule $\{\beta_t\}$
  \Ensure Trained parameterized neural network $\mathbf{z}'_\theta(\mathbf{x}_t, t)$
  \State Compute cumulative deformation coefficients: $\alpha_t = 1 - \beta_t, \quad \bar{\alpha}_t = \prod_{i=1}^t \alpha_i$
  \For{each training iteration}
    \State Randomly sample a time step $t \sim \text{Uniform}(1, T)$
    \State Sample standard Gaussian noise $\mathbf{z} \sim \mathcal{N}(0, I)$
    \State Calculate $\mathbf{Q}_t$
    \State Generate deformed geometry $\mathbf{x}_t = \sqrt{\bar{\alpha}_t}
    \mathbf{x}_0 + \sqrt{1 - \bar{\alpha}_t} \mathbf{z}+\mathbf{Q}_t$ at each timestep
    \State Using parameterized neural network $\mathbf{z}'_\theta(\mathbf{x}'_t, t)$ to predict noise
    \State Compute loss:
    $\mathcal{L} = \| \mathbf{z} - \mathbf{z}'_\theta(\mathbf{x}_t, t) \|^2$
    \State Update neural network by minimizing $\mathcal{L}$
  \EndFor
\end{algorithmic}
\end{algorithm}

\begin{algorithm}[!htbp]
\caption{Sampling from the DDPM-Polycube model}
\begin{algorithmic}[1]
  \Require Trained parameterized neural network $\mathbf{z}'_\theta(\mathbf{x}_t,
  t)$, number of diffusion steps $T$, deformation schedule $\{\beta_t\}$, input
  geometry $\mathbf{x}'_T$
  \Ensure Generated polycube structure $\mathbf{x}'_0$
  \For{$t = T, T-1, \dots, 1$}
    \State Calculate $\mathbf{q}'$
    \State Compute $\sigma_t = \sqrt{\beta_t}$,
    \[\mathbf{\mu}_{t}' = \frac{1}{\sqrt{\alpha_t}} \left( \mathbf{x}'_t - \frac{\beta_t}{\sqrt{1-\bar{\alpha}_t}} (\mathbf{z}'_\theta(\mathbf{x}'_t, t) +\mathbf{q}')\right).\]
    \State Sample deformation $\mathbf{z}_t' \sim \mathcal{N}(0, I)$ if $t > 1$, otherwise $z = 0$
    \State Update sample: $\mathbf{x}'_{t-1}=\mathbf{\mu}_{t}'+\sigma_t \mathbf{z}_t'$
  \EndFor

\Return $\mathbf{x}'_0$
\end{algorithmic}
\end{algorithm}

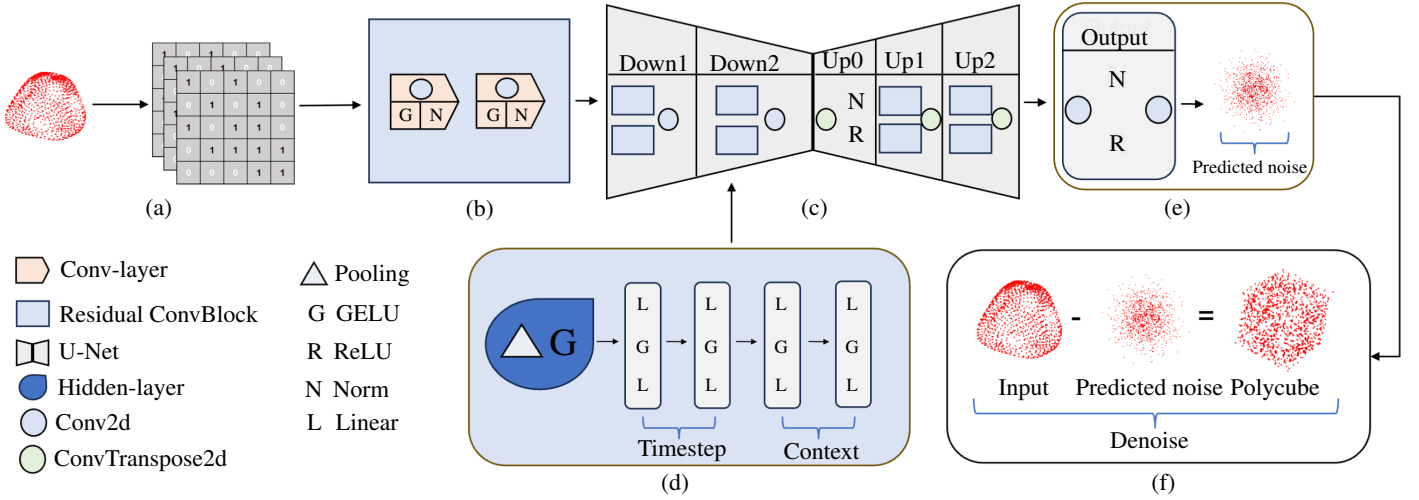
\begin{figure*}[t]
      \centering
  \begin{tikzpicture}
    \node[anchor=south west,inner sep=0] (image) at (0,0) {\includegraphics[width=\linewidth]{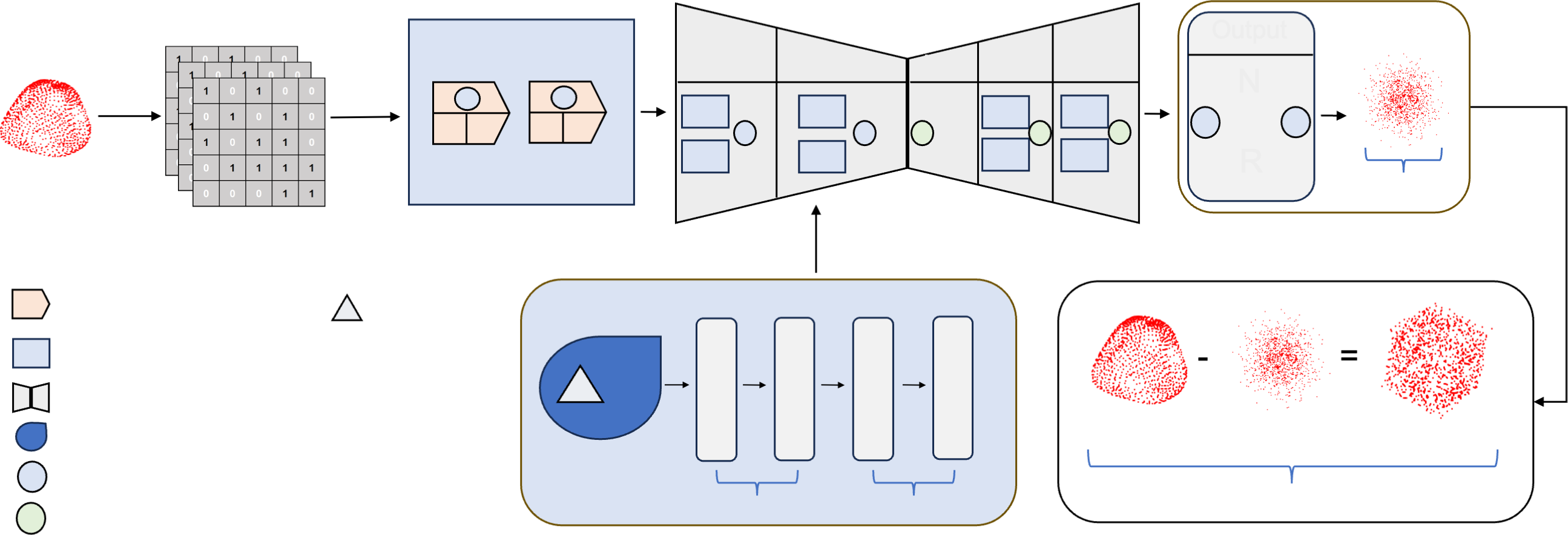}};
    \begin{scope}[x={(image.south east)},y={(image.north west)}]
      \node at (0.11,0.56) {\small (a)};
      \node at (0.34,0.56) {\small (b)};
      \node at (0.58,0.56) {\small (c)};
      \node at (0.84,0.56) {\small (e)};
      \node at (0.48,-0.02) {\small (d)};
      \node at (0.83,-0.02) {\small (f)};
      \node at (0.078,0.43) {\small Conv-layer};
      \node at (0.11,0.34) {\small Residual ConvBlock};
      \node at (0.06,0.26) {\small U-Net};
      \node at (0.083,0.18) {\small Hidden-layer};
      \node at (0.063,0.11) {\small Conv2d};
      \node at (0.097,0.03) {\small ConvTranspose2d};
      \node at (0.263,0.42) {\small Pooling};
      \node at (0.25,0.34) {\small G\hspace{4pt}GELU};
      \node at (0.247,0.26) {\small R\hspace{4pt}ReLU};
      \node at (0.245,0.18) {\small N\hspace{4pt}Norm};
      \node at (0.249,0.11) {\small L\hspace{3pt} Linear};
      \node at (0.287,0.76) {\scriptsize G};
      \node at (0.309,0.76) {\scriptsize N};
      \node at (0.35,0.76) {\scriptsize G};
      \node at (0.37,0.76) {\scriptsize N};
      \node at (0.465,0.87) {\small Down1};
      \node at (0.53,0.87) {\small Down2};
      \node at (0.60,0.87) {\small Up0};
      \node at (0.61,0.79) {\small N};
      \node at (0.61,0.72) {\small R};
      \node at (0.645,0.87) {\small Up1};
      \node at (0.695,0.87) {\small Up2};
      \node at (0.795,0.92) {\small Output};
      \node at (0.797,0.835) {\small N};
      \node at (0.797,0.7) {\small R};
      \node at (0.892,0.65) {\scriptsize Predicted noise};
      \node at (0.73,0.18) {\small Input};
      \node at (0.82,0.185) {\small Predicted noise};
      \node at (0.91,0.18) {\small Polycube};
      \node at (0.82,0.073) {\small Denoise};
      \node at (0.485,0.05) {\small Timestep};
      \node at (0.585,0.05) {\small Context};
      \node at (0.457,0.36) {\scriptsize L};
      \node at (0.457,0.27) {\scriptsize G};
      \node at (0.457,0.19) {\scriptsize L};
      \node at (0.506,0.36) {\scriptsize L};
      \node at (0.506,0.27) {\scriptsize G};
      \node at (0.506,0.19) {\scriptsize L};
      \node at (0.556,0.36) {\scriptsize L};
      \node at (0.556,0.27) {\scriptsize G};
      \node at (0.556,0.19) {\scriptsize L};
      \node at (0.607,0.36) {\scriptsize L};
      \node at (0.607,0.27) {\scriptsize G};
      \node at (0.607,0.19) {\scriptsize L};
      \node at (0.4,0.278) {\Large G};
    \end{scope}
  \end{tikzpicture}
  \vspace{-2em}\caption{\label{fig:06_markov_chain}{ Overview of the DDPM-Polycube model
      architecture. (a) Input point cloud features are preprocessed into a
      3-channel format. (b) The ResNet module within the U-Net extracts and
      refines geometric features through residual connections. (c) U-Net employs
      skip connections between downsampling and upsampling modules for detailed
      feature preservation. (d) Timestep and context embeddings are fused with
      convolutional features to enhance noise modeling. (e) The output layer
      predicts the non-standard Gaussian noise. (f) The denoising process
      iteratively removes non-standard Gaussian noise from the input geometry to
      generate the polycube structure.}}
\end{figure*}

During the forward diffusion process (see Algorithm 1), the polycube structure
data is progressively added by the small-scale deformation, which follows a
non-standard Gaussian noise.  This process follows a predefined deformation
schedule. In our paper, the schedule is designed linearly. It is controlled by
the initial and final noise values (\(\beta_1 = 1 \times 10^{-4}\),
\(\beta_T = 0.02\)). The deformation increases gradually over time.  This design
ensures that earlier timesteps retain more of the original polycube structure,
while later stages become increasingly dominated by large-scale
deformation. Specifically, for a polycube structure \(\mathbf{x}_0\), the
deformed geometry \(\mathbf{x}_t\) at timestep \(t\) is generated using
Equation~\eqref{eq:2}.

Once training is complete, the reverse diffusion process (i.e., the sampling
from the DDPM-Polycube model) iteratively deforms the input geometry to
reconstruct the polycube structure (see Algorithm 2). Starting from the final
timestep \(T\), the model predicts the deformation at each step and uses it to
compute the mean and variance of the deformation distribution. Specifically, at
timestep \(t\), the mean \(\mu'_t\) (Equation~\eqref{eq:5}) and variance
\(\sigma_t^2\) of the non-standard Gaussian noise are calculated. By sampling
from this distribution, the model progressively removes deformation at each
step, ultimately generating the polycube structure.

The architecture of the DDPM-Polycube model is built upon a U-Net framework (see
Fig.~\ref{fig:06_markov_chain}), which includes an initial convolutional layer,
two downsampling modules and two upsampling modules, as well as fully connected
embedding layers for timestep and context features. The U-Net framework
leverages skip connections to transfer features between the downsampling and
upsampling modules. This helps preserve both geometric details and the
topological structure.

Before training, the input geometry undergoes preprocessing, including
normalization, scaling, and mapping onto a three-channel $32\times 32$ matrix to
create image-like data structures compatible with the U-Net framework (see
Fig.~\ref{fig:06_markov_chain}(a)).  The initial convolutional layer (see
Fig.~\ref{fig:06_markov_chain}(b)) is implemented as a residual convolutional
block, taking three-channel geometric data as input and producing feature maps
with a dimension of 64. The residual convolutional block uses GELU activation
functions and batch normalization to enhance feature extraction while
maintaining stability during training. The downsampling module (see
Fig.~\ref{fig:06_markov_chain}(c)) consists of two residual convolutional blocks
followed by a max-pooling layer, progressively extracting local geometric
features while reducing spatial resolution. The feature dimensions increase to
128 during the downsampling module. Max-pooling reduces the spatial resolution
by half, enabling the extraction of higher-level features. At the same time, the
timestep and context features (see Fig.~\ref{fig:06_markov_chain}(d)) are
embedded using fully connected layers. The timestep embedding layer takes a
scalar input of dimension 1 and outputs a vector of dimension 128, while the
context embedding layer takes a 29-dimensional input (see
Section~\ref{sec:data-generation}) and produces an output of dimension
128. These embedded features are fused with convolutional features through
additive and multiplicative operations, enhancing the model's ability to capture
contextual information. Fully connected layers are used for embedding
generation, with GELU activation ensuring smooth transitions across
dimensions. The upsampling module (see Fig.~\ref{fig:06_markov_chain}(c))
consists of transposed convolutional layers combined with ReLU activation and
two residual convolutional blocks. The feature dimensions are reduced from 128
back to 64, with transposed convolutional layers doubling the spatial
resolution. To preserve the geometric details and topological structure, skip
connections are employed to transfer features between the corresponding layers
in the downsampling and upsampling modules. Finally, an output convolutional
layer (see Fig.~\ref{fig:06_markov_chain}(e)) maps the features back to the
three-channel data, producing the predicted noise.  This noise can be
interpreted as small-scale deformations. By gradually removing the noise from
the input geometry, the polycube structure is generated (see
Fig.~\ref{fig:06_markov_chain}(f)).

\section{Hex mesh generation and volumetric spline construction}

In this section, we describe how to generate high-quality hex meshes and
construct volumetric splines from the polycube structures produced by the
DDPM-Polycube algorithm. While the methods used here are based on existing
techniques~\cite{yu2024dlpolycubedeeplearningenhanced,yu2020hexgen}, we provide
a concise overview to maintain the completeness of the proposed pipeline. For a
detailed understanding of the algorithms and their underlying principles, we
refer readers
to~\cite{yu2020hexgen,wei17a,yu2024dlpolycubedeeplearningenhanced}.  The
implementation of these methods has been integrated into HexGen and Hex2Spline
software packages, which are publicly available on
GitHub~\cite{githexgenhex2spline}. These robust tools enable users to
efficiently generate high-quality hex meshes and construct volumetric splines
for IGA applications.

\subsection{Hex mesh generation}
To generate a high-quality hex mesh, the polycube structure and its associated
surface segmentation serve as the foundation. The process begins by establishing
a bijective mapping between the input triangular mesh and the boundary surface
of the polycube. This mapping is achieved through parametric mapping techniques,
where each segmented surface patch corresponds to a boundary surface of the
polycube. Utilizing the cotangent Laplace operator, we ensure that the
parametric mapping is harmonic and bijective. Internal regions of the polycube
are parameterized using linear interpolation. The hex mesh is constructed from
this parametric mapping in conjunction with octree subdivision, where each cubic
region of the polycube structure is subdivided into smaller cubes to generate
volumetric mesh elements. Vertices are then mapped from the parametric domain to
the physical domain. To address mesh quality issues that arise during the hex
mesh generation, we integrate several quality improvement techniques, including
pillowing, smoothing, and optimization. Pillowing improves mesh quality by
inserting additional layers around the boundary. Smoothing relocates vertices to
improve mesh quality while preserving geometric feature. The optimization
improves the scaled Jacobian~\cite{zhang2006adaptive} by minimizing an energy
function. This energy function incorporates geometry fitting and element shape
metrics. By alternately applying these optimization techniques, a high-quality
hex mesh is generated.

\subsection{Volumetric spline construction}
Following the generation of a high-quality hex mesh, the next step involves
constructing volumetric splines, specifically TH-spline3D, on the unstructured
hex mesh. The spline construction begins by using the hex mesh as the control
mesh. The TH-spline3D framework is employed to define spline functions over the
hex mesh, preserving sharp features and supporting local refinement. The
constructed volumetric splines maintain $C^0$ continuity around extraordinary
points and $C^2$ continuity in regular regions, ensuring smoothness and accuracy
for IGA simulations. To facilitate integration with simulation platforms,
Hex2Spline~\cite{githexgenhex2spline} can export the B\'{e}zier information of
the volumetric splines. This enables seamless compatibility with IGA solvers
such as ANSYS-DYNA.

\section{Results and discussion}

\subsection{Training performance of the DDPM-Polycube model}
\label{sec:model-training-ddpm}

During the forward diffusion process, non-standard Gaussian noise is added to
the polycube structures. The polycube structures include 9 types, as described
in Section~\ref{sec:data-generation}. This noise progressively corrupts the
polycube structures. In the reverse diffusion process, the model predicts and
removes the noise step by step. This restores the polycube structure. The
training objective minimizes the MSE (see Equation~\eqref{eq:4}) between the
true noise and the predicted noise at each timestep. This ensures reconstruction
of the polycube structure from the input geometry. The diffusion process is
configured with 500 timesteps. Increasing the number of timesteps can improve
reconstruction quality but also raises computational costs. Details about the
DDPM-Polycube model are provided in Section~\ref{sec:model-training}.  The model
is trained with a batch size of 200 over 400 epochs. The initial learning rate
\( \eta \) is set to $1 \times 10^{-3}$. It is adjusted using a linear decay
strategy based on the current epoch. The formula is
$\eta_k = \eta_{k-1} \times \left( 1 - \frac{k}{K} \right)$, where \( \eta_k \)
is the learning rate for the current epoch, \( k \) is the current epoch, and
\( K \) is the total number of epochs. The Adam optimizer is used to improve
training stability.

\begin{figure}[!htp]
\centering
\begin{tabular}{c}
  \includegraphics[width=0.8\linewidth]{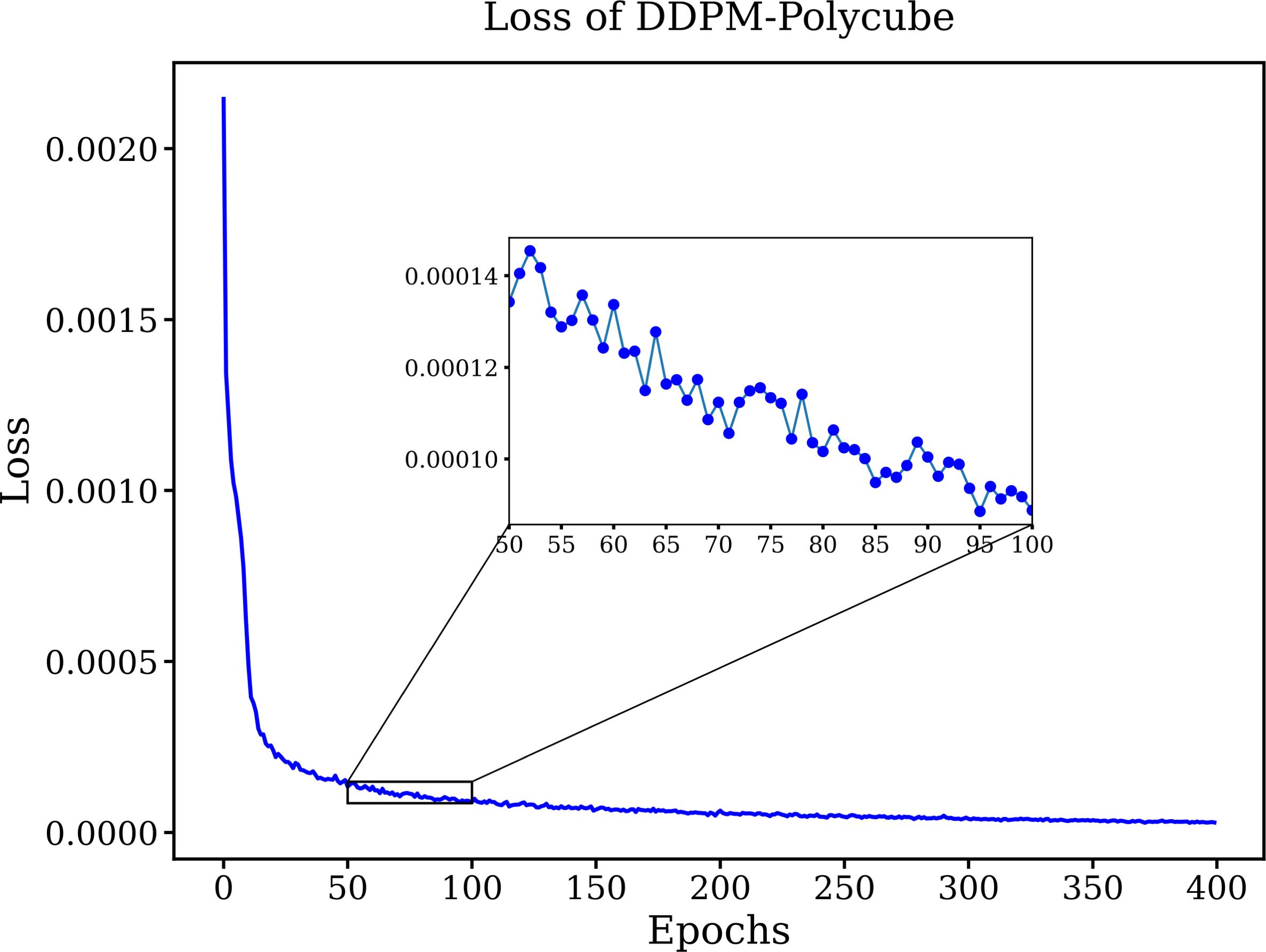}
\end{tabular}
\vspace{-1em}\caption{Loss of the DDPM-Polycube model as a function of epochs.}
\label{fig:model_performance}
\end{figure}

To assess the performance of the proposed DDPM-Polycube algorithm, we
implemented it using Python on a system equipped with an Intel Xeon CPU, 64 GB
of RAM, and an RTX 4080 GPU. Fig.~\ref{fig:model_performance} displays the loss
function of the DDPM-Polycube model as a function of epochs, demonstrating the
convergence behavior during training.

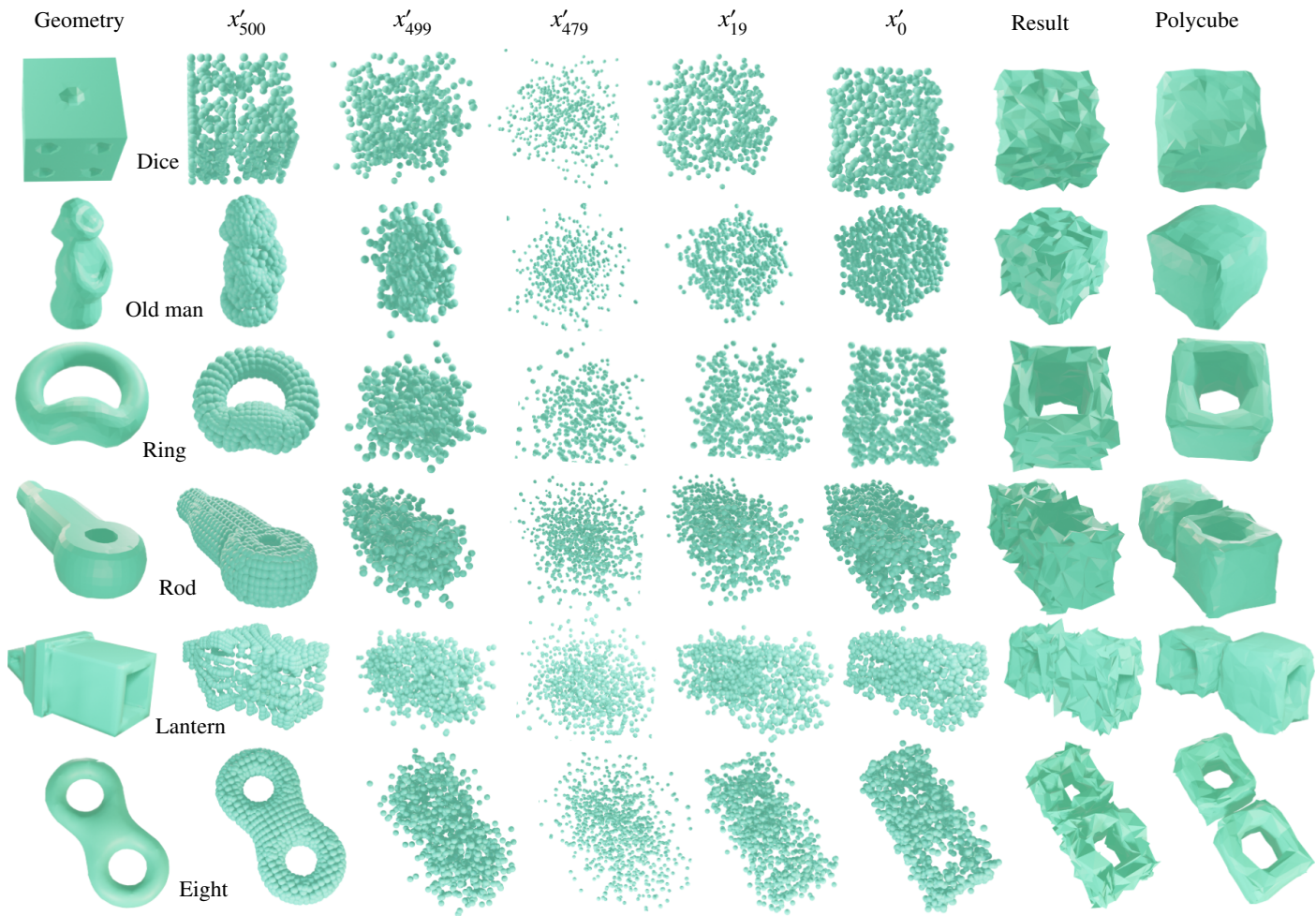
\begin{figure*}[t]
  \centering
  \begin{tikzpicture}
    \node[anchor=south west,inner sep=0] (image) at (0,0) {\includegraphics[width=\linewidth]{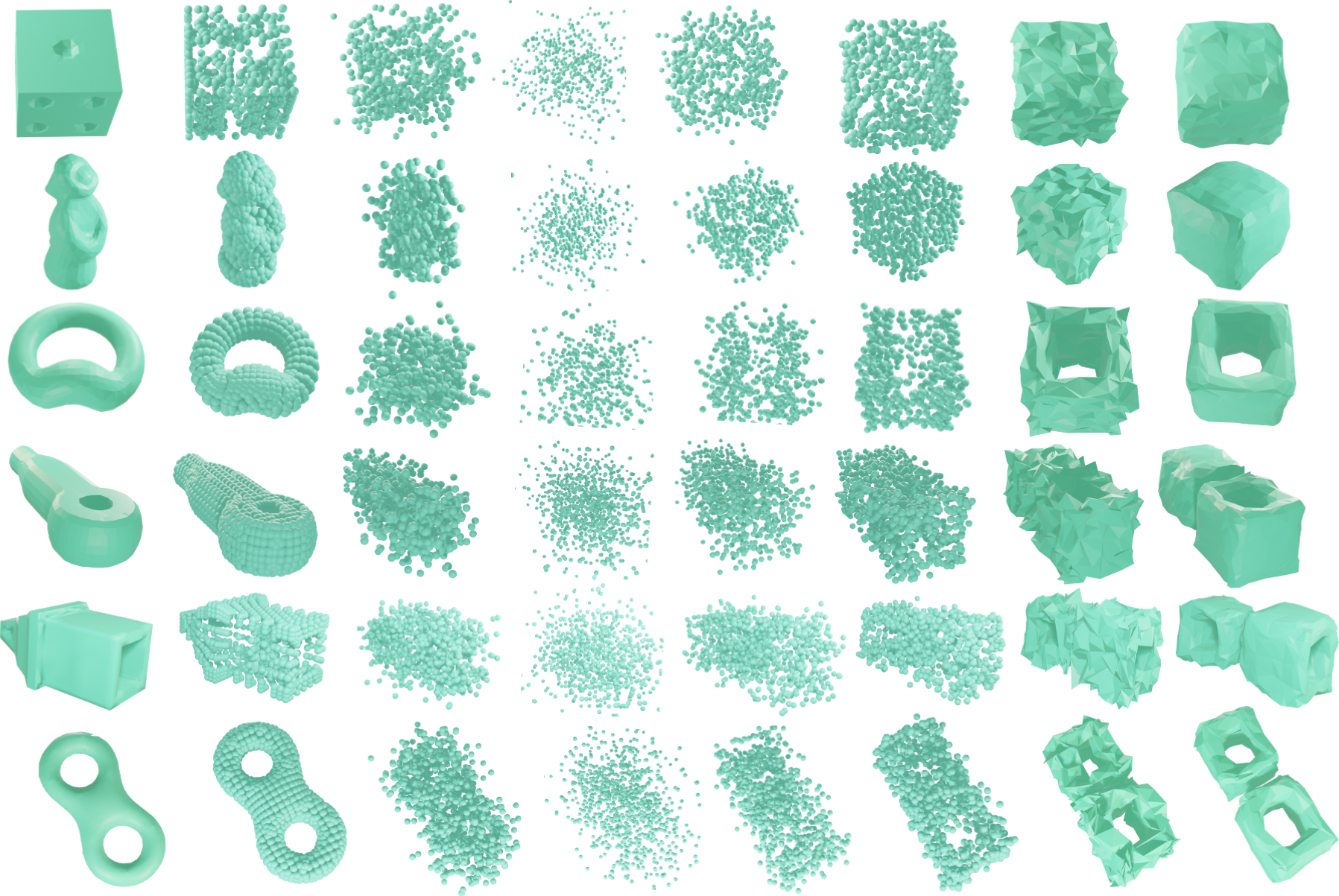}};
    \begin{scope}[x={(image.south east)},y={(image.north west)}]
      \node at (0.115,0.87) {\small Dice};
      \node at (0.12,0.7) {\small Old man};
      \node at (0.12,0.535) {\small Ring};
      \node at (0.13,0.38) {\small Rod};
      \node at (0.14,0.22) {\small Lantern};
      \node at (0.15,0.03) {\small Eight};
      \node at (0.183,1.03) {$x'_{500}$};
      \node at (0.31,1.03) {$x'_{499}$};
      \node at (0.43,1.03) {$x'_{479}$};
      \node at (0.555,1.03) {$x'_{19}$};
      \node at (0.68,1.03) {$x'_{0}$};
      \node at (0.055,1.03) {\small Geometry};
      \node at (0.79,1.03) {\small Result};
      \node at (0.91,1.03) {\small Polycube};
    \end{scope}
  \end{tikzpicture}
  \vspace{-1em}\caption{The DDPM-Polycube algorithm's capability to generate
    polycube structures for various geometric models with different genus
    levels, demonstrating its ability to adapt to complex topologies and
    generate polycube structures beyond its training set. Examples include
    genus-0 models (Dice, Old man), genus-1 models (Ring, Rod, Lantern), and
    genus-2 models (Eight). The reverse diffusion process starts from the input
    geometry (\(x'_{500}\)), progresses through latent variable models (\(x'_{499}\),
    \(x'_{479}\), \(x'_{19}\)), and gradually removes the noise to reconstruct
    the final polycube structure (\(x'_{0}\)). }
  \label{fig:model_genus_1}
\end{figure*}

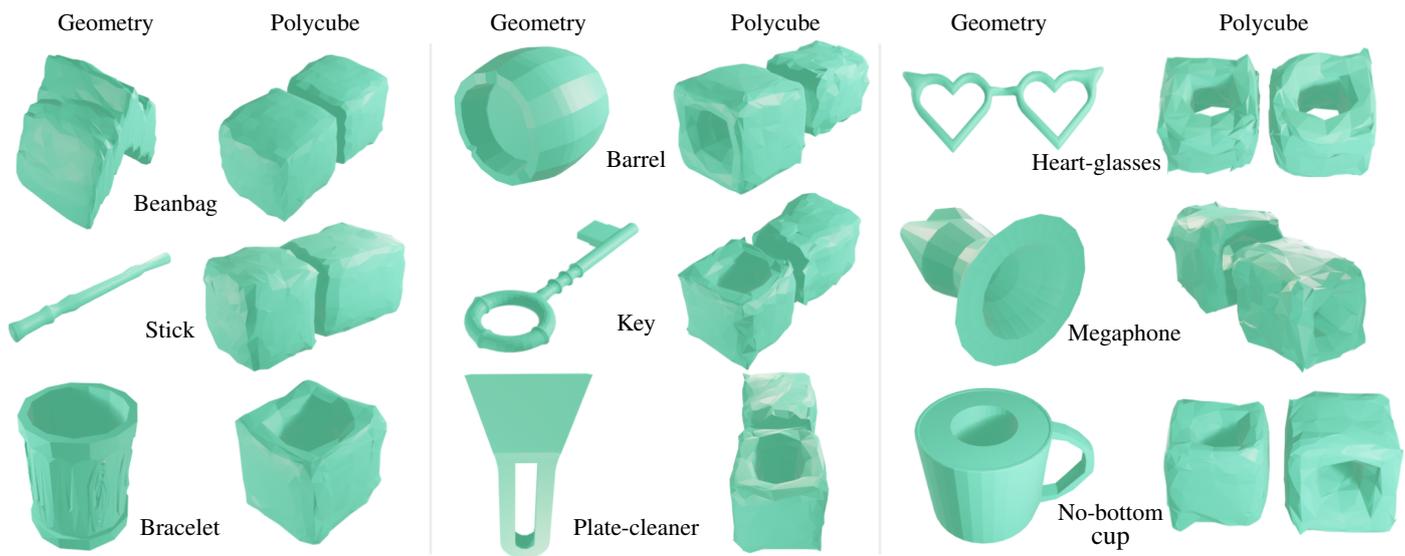
\begin{figure*}[!htp]
  \centering
  \begin{tikzpicture}
    \node[anchor=south west,inner sep=0] (image) at (0,0) {\includegraphics[width=\linewidth]{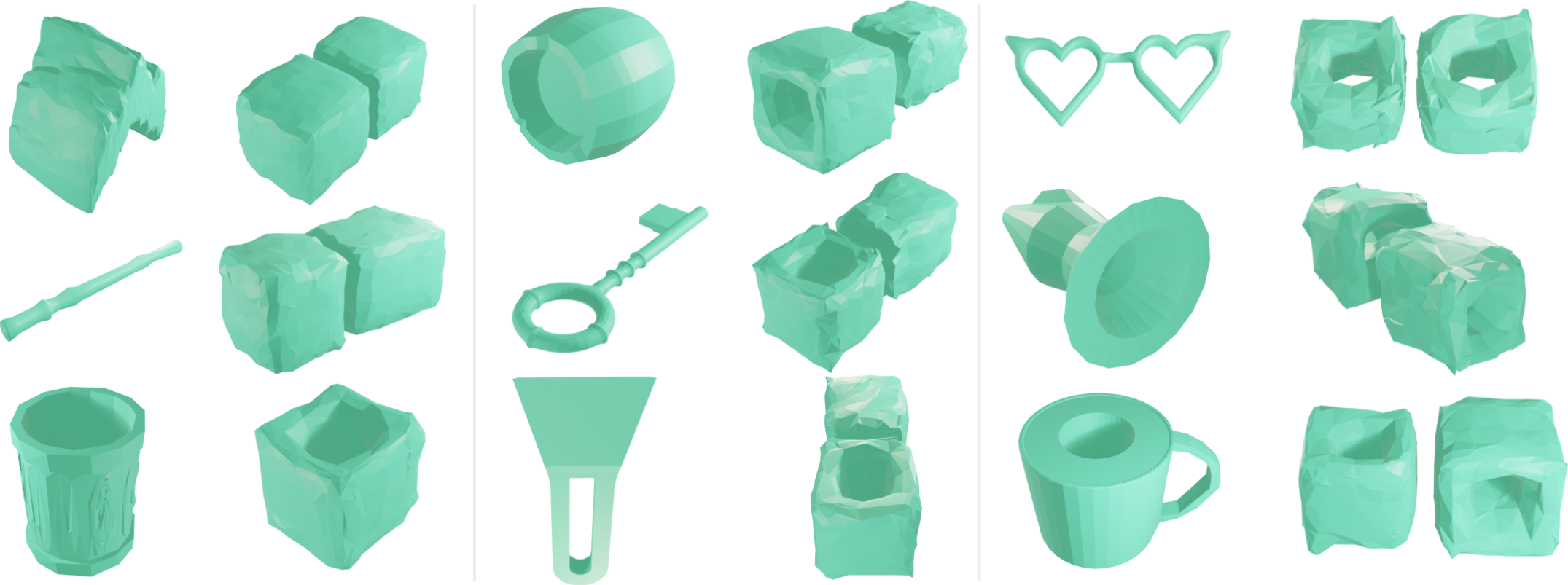}};
    \begin{scope}[x={(image.south east)},y={(image.north west)}]
      \node at (0.12,0.68) {\small{Beanbag}};
      \node at (0.116,0.44) {\small{Stick}};
      \node at (0.123,0.06) {\small{Bracelet}};
      \node at (0.45,0.77) {\small{Barrel}};
      \node at (0.45,0.45) {\small{Key}};
      \node at (0.45,0.06) {\small{Plate-cleaner}};
      \node at (0.78,0.76) {\small{Heart-glasses}};
      \node at (0.8,0.43) {\small{Megaphone}};
      \node at (0.79,0.06) [align=center] {\small  No-bottom\\[-0.3em]cup};
      \node at (0.07,1.03) {\small{Geometry}};
      \node at (0.22,1.03) {\small{Polycube}};
      \node at (0.38,1.03) {\small{Geometry}};
      \node at (0.55,1.03) {\small{Polycube}};
      \node at (0.71,1.03) {\small{Geometry}};
      \node at (0.9,1.03) {\small{Polycube}};
    \end{scope}
  \end{tikzpicture}
  \vspace{-1em}\caption{Additional results demonstrating the capability of the
    DDPM-Polycube algorithm in handling a wide range of geometric models. This
    includes examples of various genus-0 (Beanbag, Stick, Barrel), genus-1
    (Bracelet, Key, Plate-cleaner, Megaphone), and genus-2 (Heart-glasses,
    No-bottom cup) structures.}
  \label{fig:model_genus_2}
\end{figure*}

\subsection{Performance evaluation in geometries of different genus}
\label{sec:perf-eval-geom}

In this section, we evaluate the performance of the DDPM-Polycube algorithm on
models with varying genus levels. The results demonstrate its ability to
generate polycube structures, adapt to complex topologies, and recognize
geometric topologies beyond its training range (see
Figs.~\ref{fig:model_genus_1} and~\ref{fig:model_genus_2}).

Our initial tests focus on simple genus-0 and genus-1 models. Although these
geometries are not part of the training set, their topology falls within the 9
geometric configuration types shown in Fig.~\ref{fig:05_context}. For example,
the Old Man and Dice models (see Fig.~\ref{fig:model_genus_1}) are reconstructed
as Type I polycube structures. These models are represented as single cubes. For
more complex genus-0 geometries, such as the Stick and Beanbag models (see
Fig.~\ref{fig:model_genus_2}), the algorithm generates Type IX polycube
structures. These structures consist of two cubes. For genus-1 models, the
algorithm remains effective. The Ring model (see Fig.~\ref{fig:model_genus_1})
and the Bracelet model (see Fig.~\ref{fig:model_genus_2}) are reconstructed as Type
III polycube structures (see Fig.~\ref{fig:05_context}). The results confirm
that the algorithm captures the deformation required to represent their polycube
structures. In summary, the DDPM-Polycube algorithm successfully generates
polycube structures for models with topologies similar to those from Type I to
Type IX.

Now, we turn our attention to geometries whose topology falls outside the
training range. For the Barrel model (see Fig.~\ref{fig:model_genus_2}), which
is genus-0, there are no similar topologies in the dataset. Despite this, the
algorithm successfully captures the deformation and generates a valid polycube
structure. The barrel falls under the category of punctured geometries. In these
geometries, the presence of holes does not alter their fundamental topological
properties. To match the topology of the barrel, the algorithm generates one
cube and one cube with a hole. In this configuration, one cube’s wall covers the
hole of the other. This results in a polycube structure consistent with the
barrel’s topology. These results demonstrate the DDPM-Polycube algorithm’s
flexibility in handling punctured geometries and producing valid polycube
structures.

For genus-1 models, such as the Rod (see Fig.~\ref{fig:model_genus_1}) and Key
(see Fig.~\ref{fig:model_genus_2}) models, the algorithm successfully captures
their topological features, even though no similar topologies exist in the
training dataset. It deforms the upper part of these models into a cube and the
lower part into a cube with a hole. Similarly, for the Lantern model (see
Fig.~\ref{fig:model_genus_1}), the algorithm handles the additional complexity
of punctured geometries, generating a polycube structure where the wall of one
cube with a hole covers another cube with a hole, while preserving the overall
geometric topology. These results highlight the algorithm’s ability to recognize
and process punctured geometries with genus-1 features. For genus-2 models,
which present greater complexity, the DDPM-Polycube algorithm continues to
perform effectively. In cases such as the Eight model (see
Fig.~\ref{fig:model_genus_1}) and Heart-glasses model (see
Fig.~\ref{fig:model_genus_2}), the algorithm successfully identifies the two
holes and generates polycube structures where both the upper and lower parts
feature holes aligned in the same direction.

We also test other complex genus-1 and genus-2 geometries to further evaluate
the generalization capability of the algorithm. For the Plate-cleaner model (see
Fig.~\ref{fig:model_genus_2}), the algorithm generates a polycube structure
where the upper part deforms into a cube and the lower part into a cube with a
hole. For the Megaphone model (see Fig.~\ref{fig:model_genus_2}), it identifies
the cubes with holes in both the upper and lower parts, aligned in the same
direction. In the No-bottom cup model (see Fig.~\ref{fig:model_genus_2}),
which has genus-2 topology, the algorithm generates two genus-1 cubes with
perpendicular axes.

Overall, although our dataset only includes models from Type I to Type IX, the
experimental results demonstrate that the DDPM-Polycube algorithm can generate
polycube structures across a variety of complex geometric models, even
recognizing topologies beyond its training range. The algorithm's ability to
handle models with different genus levels and topological geometry without
predefined polycube templates highlights its robustness and versatility. This
makes it a valuable tool for polycube-based hexahedral mesh generation.

\begin{remark}[Reasoning behind the model's effectiveness]
  The model effectively uses the learned deformation capabilities of two
  geometric primitives: a cube and a cube with a hole. It identifies regions in
  the input geometry's point cloud that resemble these primitives in
  topology. These regions are then deformed into the corresponding
  structures. For example, cube-like areas are deformed into cube structures,
  while areas resembling a cube with a hole are deformed into cube with hole
  structures. By combining these learned geometric structures, the model
  generates new polycube structures, including complex CAD geometries with
  diverse topologies beyond its training range. This highlights the model's
  ability to process and generalize to complex geometric structures.
\end{remark}

\begin{figure*}[t]
  \centering
  \begin{tikzpicture}
    \node[anchor=south west,inner sep=0] (image) at (0,0) {\includegraphics[width=\linewidth]{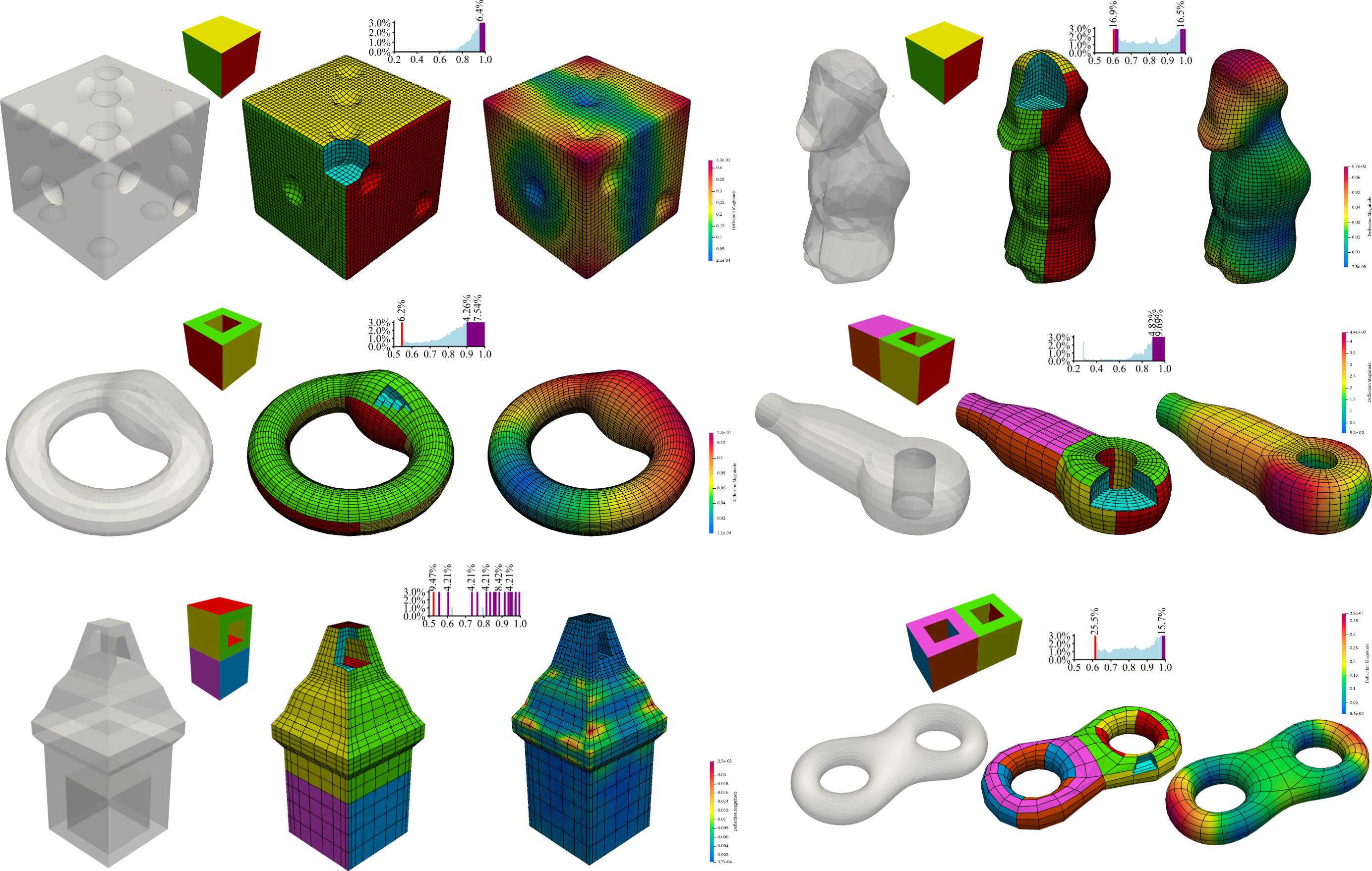}};
    \begin{scope}[x={(image.south east)},y={(image.north west)}]
      \node at (0.07,0.96) {\small Dice};
      \node at (0.55,0.95) {\small Old man};
      \node at (0.07,0.6) {\small Ring};
      \node at (0.55,0.575) {\small Rod};
      \node at (0.06,0.32) {\small Lantern};
      \node at (0.58,0.17) {\small Eight};
    \end{scope}
  \end{tikzpicture}
  \vspace{-1em}\caption{Results of the Dice, Old man, Ring, Rod, Lantern, and Eight models,
    including their polycube structures, all-hex control meshes (some elements
    are removed to show the interior), scaled Jacobian histograms, and
    volumetric splines with IGA eigenvalue analysis in ANSYS-DYNA. The red bar
    in the histograms represents the minimum scaled Jacobian.  The number of
    vertices and the number of elements for these six models are (35,937,
    32,768), (23,395, 20,992), (22,592, 20,480), (4,520, 3,840),
    (3,348, 2,565), and (1,253, 992), respectively.}
  \label{fig:hex_spline}
\end{figure*}
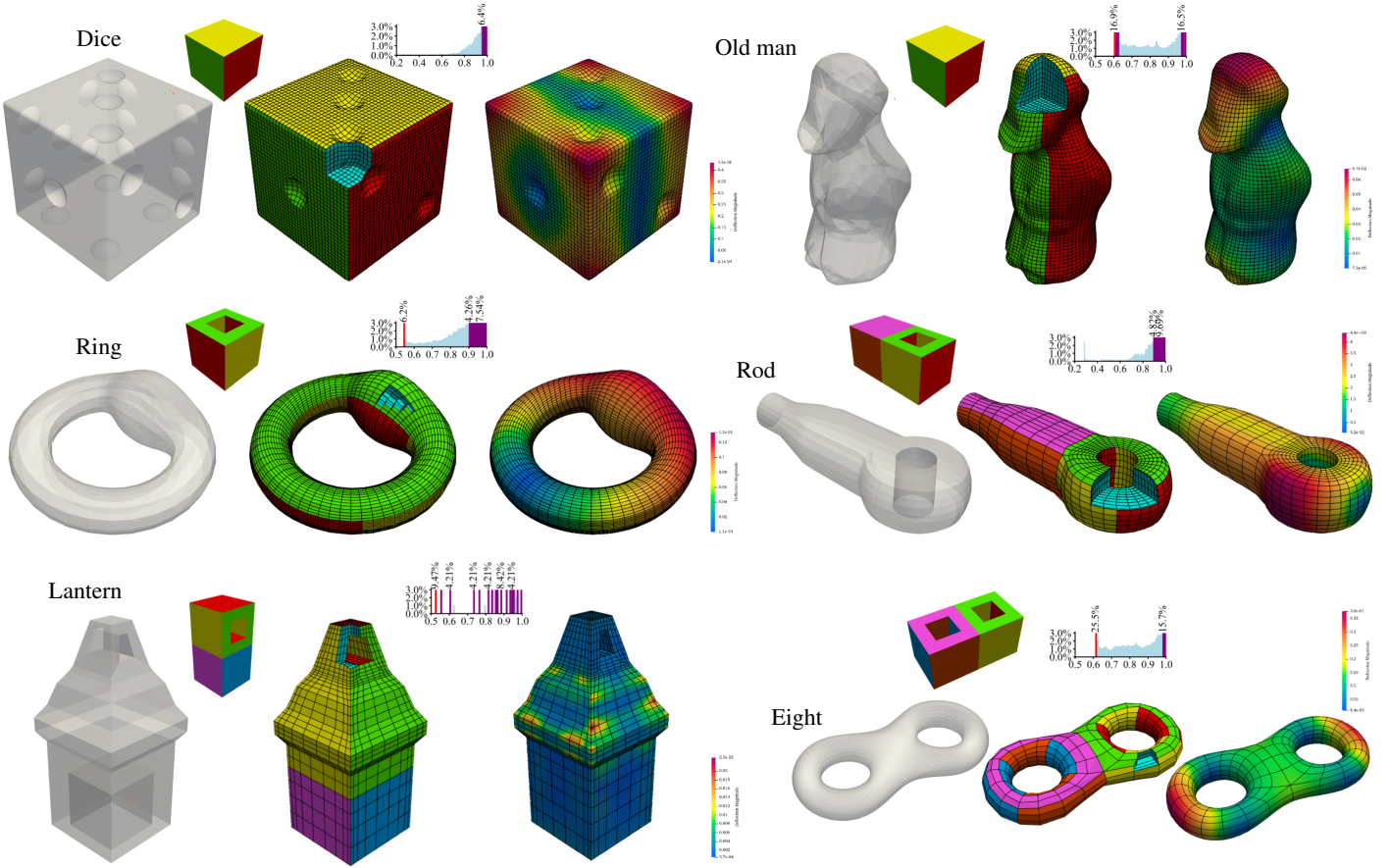

\subsection{Hex mesh generation and volumetric spline construction}
\label{sec:hexah-mesh-gener}

After constructing the polycube structure, the next step is to generate hex
meshes through surface segmentation and parametric mapping. Specifically, the
surface segmentation ensures that the segmented regions align with the polycube
structure, enabling a one-to-one correspondence between the surface of the CAD
geometry and the surface of the polycube structure. This method, which performs
segmentation after constructing the polycube structure, resolves the issues of
traditional polycube methods, where segmentation or labeling often fails to
ensure valid polycube structures. Since the segmentation or labeling in our
method is inherently based on the polycube structure, it naturally guarantees
the validity of the polycube structure. With the polycube structure serving as
the parameter space, hex mesh generation is achieved through parametric mapping
and octree subdivision. To further improve the quality of the hex mesh,
post-processing techniques such as smoothing, pillowing, and optimization are
applied. These techniques enhance the aspect ratio, reduce element distortion,
and ensure that the mesh quality meets the requirements for IGA.
Fig.~\ref{fig:hex_spline} shows the polycube structures, all-hex meshes, and
scaled Jacobian histograms for the Dice, Old Man, Ring, Rod, Lantern, and Eight
models. The obtained all-hex meshes exhibit good quality (minimal Jacobian
$> 0.27$).

Once a high-quality hex mesh is obtained, we test these models for IGA using
TH-spline3D. The results of the spline construction exhibit \(C^0\)-continuity
around extraordinary points and edges, while maintaining \(C^2\)-continuity in
all other regions. Subsequently, B\'{e}zier elements are extracted for IGA
analysis. For each tested model, we used ANSYS-DYNA to perform eigenvalue
analysis and present the results of one mode (see
Fig.~\ref{fig:hex_spline}).

\section{Conclusions and future work}

This paper proposes a novel diffusion-based deep learning method, DDPM-Polycube,
for generating high-quality hex meshes and constructing volumetric splines. By
utilizing the improved DDPM, our method models the deformation from input
geometry to polycube structures as a denoising task. This method eliminates the
dependency of DL-Polycube algorithms on predefined polycube templates, enabling
better generalization to complex geometries and diverse topologies. Experimental
results demonstrate that DDPM-Polycube can generate valid polycube structures
for complex engineering geometries and further use them to construct
high-quality hex meshes and volumetric splines that meet the requirements of
IGA. Additionally, the integration of DDPM-Polycube with mesh quality
improvement techniques and TH-spline3D further enhances the automation of the
process from B-Rep geometry to volumetric parameterization.

While the DDPM-Polycube algorithm represents an improvement on hex mesh
generation and volumetric spline construction, there are several areas for
future improvement. First, the current algorithm primarily focuses on simple
genus-0, genus-1, and genus-2 geometries. Extending the model to handle
higher-genus and more complex geometries is an interesting research
direction. Second, while the method can generate valid polycube structures,
there is still room for improvement in deformation modeling and noise
parameterization. Since the generated polycube structures carry non-standard
Gaussian noise, we adopt volume-preserving Laplacian smoothing, a feature
provided by the 3D graphics software Blender. However, for complex geometries,
more efficient algorithms are needed to transform polycube structures with
Gaussian noise into regular polycube structures. Third, the dataset used to
train the DDPM-Polycube model includes only a limited set of geometric
primitives. Expanding the dataset to incorporate more types of geometric
primitives could significantly improve the model's robustness and generalization
capability. Fourth, we observe during research that the algorithm has the
ability to decompose geometries into geometric primitives. This feature has
potential applications in another branch of IGA, namely volumetric
parameterization based on CSG. Fifth, the application of context in this method
is particularly important. If geometric information, connectivity relationships,
or even basic data such as STL files could replace the current context, it would
bring the method closer to full automation. Furthermore, combining more
multimodal learning frameworks could also further enhance the performance of
DDPM-Polycube. Finally, although this study primarily focuses on hex mesh
generation, applying diffusion-based generative models to other types of meshes
(such as hybrid meshes or hex-dominant meshes) could broaden the scope of this
method. Moreover, extending the framework to support direct integration with CAD
environments and automated IGA workflows could significantly improve its
practical usability in real-world engineering applications. By addressing these
challenges, we believe that DDPM-Polycube has the potential to advance the
fields of computational geometry, mesh generation, and IGA.

% \bmhead{Acknowledgements}

% This project is supported by the National Natural Science Foundation of China
% (Grant No. 62302091), the Discipline Innovation Field Cultivation Project (Grant
% No. XKCX202308), and the Fundamental Research Funds for the Central Universities
% (Grant No. 2232023D-24). H. Tong and Y. J. Zhang were supported in part by a
% Honda project.

%%
\bibliographystyle{elsarticle-num}
\bibliography{Untitled.bib}



\end{document}